\documentclass[fleqn,usenatbib]{mnras}

\usepackage{newtxtext,newtxmath}

\usepackage[T1]{fontenc}
\usepackage{ae,aecompl,times}

\usepackage{graphicx}	
\usepackage{amsmath}	
\usepackage{amssymb}	

\def\Mpch{~h^{-1}{\rm Mpc}}

\newcommand{\revise}[1]{\textcolor{black}{#1}}

\title[Iterative RSD reconstruction]{Iterative removal of redshift space distortions from galaxy clustering}

\author[Wang et al.]{
Yuchan Wang,$^{1}$\thanks{E-mail: yuchan.wang@durham.ac.uk}
Baojiu Li$^{2}$
and Marius Cautun$^{3}$
\\
$^{1}$Department of Physics, Durham University, South Road, Durham DH1 3LE, United Kingdom\\
$^{2}$Institute for Computational Cosmology, Department of Physics, Durham University, South Road, Durham DH1 3LE, United Kingdom\\
$^{3}$Leiden Observatory, Leiden University, PO Box 9513, NL-2300 RA Leiden, the Netherlands
}

\date{Accepted XXX. Received YYY; in original form ZZZ}

\pubyear{2019}

\begin{document}
\label{firstpage}
\pagerange{\pageref{firstpage}--\pageref{lastpage}}
\maketitle

\begin{abstract}
Observations of galaxy clustering are made in redshift space, which results in distortions to the underlying isotropic distribution of galaxies. These redshift-space distortions (RSD) not only degrade important features of the matter density field, such as the baryonic acoustic oscillation (BAO) peaks, but also pose challenges for the theoretical modelling of observational probes. Here we introduce an iterative nonlinear reconstruction algorithm to remove RSD effects from galaxy clustering measurements, and assess its performance by using mock galaxy catalogues. The new method is found to be able to recover the real-space galaxy correlation function with an accuracy of $\sim1\%$, and restore the quadrupole accurately to $0$, on scales $s\gtrsim20\Mpch$. It also leads to an improvement in the reconstruction of the initial density field, which could help to accurately locate the BAO peaks. An `internal calibration' scheme is proposed to determine the values of cosmological parameters as a part of the reconstruction process, and possibilities to break parameter degeneracies are discussed. RSD reconstruction can offer a potential way to simultaneously extract the cosmological parameters, initial density field, real-space galaxy positions and large-scale peculiar velocity field (of the real Universe), making it an alternative to standard perturbative approaches in galaxy clustering analysis, bypassing the need for RSD modelling.
\end{abstract}




\section{Introduction}


The observed large-scale cosmic structures today encode information about the primordial matter density field -- the earliest memory of our own Universe, that came from a time when the Universe was in 
a simpler form, where density perturbations can be described by linear perturbation theory and the nonlinear structure formation had not made the picture more complicated. As an example, the nearly Gaussian curvature fluctuations, as supported by observations \citep[][]{Ade:2013ydc,Ade:2015ava,Akrami:2019izv}, can teach us a lot about what has happened during inflation. The observed Universe today, however, can look very different from its initial conditions, due largely to the growth of tiny density perturbations by gravitational instability to form large, nonlinear, dark matter clumps in which galaxies, stars and planets 
evolve. Inevitable in this process is the permanent loss of certain details of the primordial state of the Universe, but it still possible to retrieve the remaining useful information by `reconstructing' the initial condition. The latter is a topic which has been investigated for several decades, with increasing interest in recent years \citep[see, e.g.,][and references therein]{Peebles1989ApJ...344L..53P,Croft1997,Brenier2003,Eisenstein2005,Zhu2017,Zhu:2017vtj,Schmittfull2017,Shi:2017gqs,Hada:2018ziy,Hada:2018fde,Birkin:2018nag,Bos:2018rpw,Wang_2019,Yu2019Vel,Zhu2019,Kitaura:2019ber}. 

One of the main motivations of initial density reconstruction is related to the extraction of the baryonic acoustic oscillation (BAO) signal from galaxy surveys. BAO is a cosmological relic of the random density fluctuations that propagated in the primordial photon-electron-nuclei plasma before recombination. At the epoch of recombination, the disappearance of free electrons stopped this propagation, so that the perturbations and their interference were frozen, leaving an imprint in the matter distribution that is detectable at late times in the galaxy distribution \citep{Eisenstein1998}. 

This imprint is a typical length scale corresponding to the sound horizon, the largest distance sound waves in the plasma could have travelled by a given time, at recombination. For this reason, BAO serves as valuable standard ruler that can be used to study the expansion history of the Universe. 

Precise measurements of cosmological distances using BAO can improve the prospective of constraining cosmological models and shedding light on the mystery of the cosmic acceleration \citep{Weinberg2013},
with forthcoming galaxy surveys \citep{SKA,DESI,EUCLID}.

However, the BAO peaks found through the observed galaxy correlation function and power spectrum are shifted, weakened and broadened \citep{Eisenstein2007,Crocce2008} by the process of nonlinear gravitational evolution and bulk motions of matter \citep{Obuljen2017}, making it harder to accurately determine the peak positions and to use them to measure cosmological distances. This is further complicated by the fact that galaxies are biased tracers of the large-scale structure, and by redshift space distortions (RSD), a phenomenon that arises because we measure the redshifts, rather than real distances, of galaxies, and the former can be affected by the large-scale peculiar velocity field, leading to incorrect interpreted galaxy coordinates. Both of the latter effects can further degrade the potential of BAO as a standard ruler \citep{Birkin:2018nag,Zhu2017}. The idea is that with reconstruction we can at least partially remove these effects, therefore improving the accuracy of cosmological constraints. 

A variety of previous reconstruction methods have found success in reducing of the effects of cosmic structure formation in the recovery of the BAO peaks. Starting from the first attempt (which is now called standard reconstruction) reversing the motion of galaxies \citep{Eisenstein2007}, which has been proved to be effective in observations \citep{Padmanabhan2012}, improvement has been found in methods using iterations \citep{Schmittfull2017}. Inspired by Lagrangian perturbation theory, which uniquely maps the final Eulerian coordinates of galaxies to a set of initial Lagrangian positions, recent developments propose that the process of reconstruction can be treated as solving an optimal mass assignment problem \citep{Frisch2002, Brenier2003, Mohayaee2006}. This problem has been lately solved as a nonlinear partial differential equation using different algorithms \citep{Zhu2017,Shi:2017gqs}. Forward-modelling reconstruction methods are also studied extensively \citep[e.g.][]{Kitaura2008,Jasche2013,Wang2014,Lavaux2016}, where efficient Monte Carlo samples of the initial density field phases are combined with nonlinear evolution to select the initial condition that would match well late-time observations of the local Universe.

The reconstruction method proposed by \citet{Shi:2017gqs} is the starting point of the iterative reconstruction scheme to be described in this work. This method reduces the reconstruction problem into solving a Monge-Ampere-type partial differential equation (PDE), which gives the mapping between the initial, Lagrangian, and final, Eulerian, coordinates of particles. In $3$ spatial dimensions, the PDE contains up to cubic powers of second-order derivatives, and can be solved using a slightly modified multigrid relaxation technique. Although originally developed for reconstructions from a dark matter field, its generalisation for reconstructions from biased tracers, such as galaxies and dark matter haloes, turned out to be straightforward \citep{Birkin:2018nag}. In this work, we will further extend this method for reconstructions from biased tracers in {\it redshift} space, by making use of the relation between the displacement field and the peculiar velocity field.

As mentioned above, RSD means that the inferred galaxy coordinate is different from its true coordinate. There are two regimes of the RSD effect, as can be illustrated by considering two galaxies, both along the line of sight (LOS), one in front of and the other behind a galaxy cluster which is along the same LOS. If these galaxies are distant from the central cluster, they fall toward the latter but the infall velocity is generally not very high -- the galaxy in front of the cluster experiences an additional redshift due to the infall velocity, making it appear further away from us, while the one behind has an additional blueshift which makes it appear to be closer to observer than its true distance. In this regime, the two galaxies would appear closer to each other, leading to a squashing (Kaiser) effect along the LOS in the galaxy correlation function. On the other hand, if the two galaxies are both much closer to the cluster centre, their velocities are likely much larger; the one in front could appear to be behind the cluster and vice version, which causes a strongly elongated feature along the LOS in the galaxy correlation function, known as the finger-of-God (FoG) effect. The large-scale Kaiser effect can be well described by linear perturbation theory, while the FoG effect, being on small scales, is nonlinear. The FoG effect causes `trajectory crossing', i.e., it changes the ranking order of the distances of galaxies, and in general this poses a limitation on reconstruction as we will discuss later. 

Assuming no trajectory crossing and a curl-free velocity field, the peculiar velocity field induced gravitationally by overdensities can be derived from the density field itself in the real space. Intuitively, the RSD effect can be described as a ``more evolved matter field'' \citep{Taylor1993}, recognising this intimate relationship between RSD and gravitational process. Accordingly, in the standard reconstruction approach, the RSD effect has been considered as an additional linear factor on the displacements of galaxies following Kaiser's equation which links the displacement field to the compression effect due to galaxy coherent motion (but neglects the FoG effect). 

Obtaining velocity field from density field through nonlinear reconstruction has been explored by \citet{Yu2019Vel}. Their result suggests that the correlation between the matter density field and the velocity field can be more complicated than the linear theory prediction. Since the nonlinear displacement field can be obtained from new reconstruction methods, including that of \citet{Shi:2017gqs}, we are interested to infer the peculiar velocity from it and subsequently use this information to ``undo'' the RSD effect on measured galaxy coordinates.

However, estimating velocities from a density field in redshift space is an inverse problem -- no real space density field is known {\it a priori} in practice. A reliable way to approach the problem could be to use an iterative approach similar to self-calibration between the real- and redshift-space density fields until one obtains a converged result. It was proposed by \citet{Yahil1991} and \citet{Strauss1989} that an iteration scheme can be used to recover the density field in real space from observations. In the linear regime, N-body simulation results confirmed the potential of this method \citep[][]{DSY1991}. However, nonlinear effect caused by the random motions of galaxies can lead to erroneous estimations, especially in high-density clusters. This can be mitigated by a smoothing of the velocity field, echoing the result found by \citet{Cole1994} where the smoothed field gave a significantly more accurate estimation of redshift distortion parameter, $\beta$. A second-order improvement of the method was proposed by \citet{Gramann1994}, and a quasi-non-linear treatment by \citet{Taylor1993}. They both found  a strong correlation with the true density field of the density reconstructed from redshift space. More recently, iterative constructions of the initial density field have been proposed by \citet{Hada:2018fde,Hada:2018ziy}, extending the work of \citet{Monaco1999}, and by \citet{Burden:2015pfa}. Our approach follows a similar iterative procedure as these more recent works, but has a number of differences. For examples, instead of reconstructing the initial density field, we aim to reconstruct the galaxy coordinates in real space because we are more interested in the removal of RSD effects from real observational data; our displacement field is obtained from nonlinear reconstruction; and we have defined different estimators (mainly in configuration space) to quantitatively examine the reconstruction results during iterations. We are interested in reconstruction in an {\it internal-calibration} sense, namely the physical and technical parameters used for reconstruction are tuned by inspecting the reconstruction outcome itself.

The paper is organised in the following way: in Section \ref{sect:method} we describe our methodology: in Section \ref{subsect:r-space-recon} we introduce the basics of the reconstruction method proposed in \citet{Shi:2017gqs}; in Section \ref{subsect:z-space-recon} we relate the displacement field to the peculiar velocity field, arguing that this link enables an iterative method in which, starting with some rough initial guess of these fields, we can gradually improve our knowledge of them during each iteration; in Section \ref{subsect:implementation} describe in great details how the method is implemented in practice and define four estimators to assess its performance. Because of the large number of symbols used in this paper, we summarise them in Table \ref{table:params} to aim the reader. In Section \ref{sect:tests}, we test the effect of choosing different physical and technical parameters in our pipeline on the reconstruction result and performance; this section is technical and readers who are more interested in the results can skip it. Section \ref{sect:application} is the main result of this paper, where we show an application of the new method, in which we use mock galaxy catalogues constructed from a suite of N-body simulations to assess the potential of using this method to simultaneously obtain the real-space galaxy coordinates, the real-space initial matter density field and determine the physical parameters of the cosmological model. Finally, we summerise the main results, discuss the outlook and future applications of the method, and conclude in Section \ref{sect:conclusions}. 

The main figures of this paper are Figure \ref{fig:flowchart} (schematic description of the method) and Figures \ref{fig:15_combines}, \ref{fig:Beta_result} (performance illustration).

\section{Methodology}
\label{sect:method}

\subsection{Nonlinear reconstruction in real space}
\label{subsect:r-space-recon}

The iterative RSD reconstruction method described in this paper is based on the real-space nonlinear reconstruction method introduced by \citet{Shi:2017gqs}; see also \citet[][]{Baojiu2018book}. For completeness, here we briefly recap the basic idea behind that method.

Our main objective is to identify a mapping between the initial Lagrangian coordinate, ${\bf q}$, of a particle and its Eulerian coordinate, ${\bf x}(t)$, at some later time $t$. Such a mapping can be uniquely obtained, at least under the condition that the trajectories of particles have not crossed each other, by starting from the following equation,
\begin{equation}\label{eq:mass_conservation}
\rho({\bf x}){\rm d}^{3}{\bf x} = \rho({\bf q}){\rm d}^{3}{\bf q} \approx \bar{\rho}{\rm d}^{3}{\bf q},
\end{equation}
which is based on continuity equation stating that mass is conserved in an infinitesimal volume element. $\rho({\bf q})$ and $\rho({\bf x})$ are, respectively, the initial density field and the density field at time $t$. 
As the density field is very close to homogeneous at early times, we can approximate the initial  $\rho({\bf q})$ as a constant, $\rho({\bf q})\simeq\bar{\rho}$.

The displacement field, ${\bf\Psi}({\bf x})={\bf x}-{\bf q}$, between the final and initial positions of a particle can be rewritten as
\begin{equation}\label{eq:Theta_defination}
    {\bf\nabla}_{\bf x}{\Theta}({\bf x}) \equiv {\bf q} = {\bf x} - {\bf\Psi}({\bf x}),
\end{equation}
where $\Theta({\bf x})$ is the displacement potential, whose gradient is ${\bf q}$. Underlying these definitions is another approximation in this method, namely the displacement field is curl-free,  ${\bf\nabla}\times{\bf\Psi}=0$, which should break down on small scales. Substituting Eq.~(\ref{eq:Theta_defination}) into Eq.~(\ref{eq:mass_conservation}), we get
\begin{equation}\label{eq:jacobian}
    \det[\nabla^{i}\nabla_{j}\Theta({\bf x})] = \frac{\rho({\bf x})}{\bar{\rho}} \equiv 1 + \delta({\bf x}),
\end{equation}
where  $i,j$ runs over $1,2,3$ and $\delta({\bf x})$ is the density contrast at time $t$. The symbol `${\det}$' denotes the determinant of a matrix, in this case the Hessian of $\Theta({\bf x})$. A new algorithm to solve Eq.~\eqref{eq:jacobian} was developed in \citet{Shi:2017gqs}, which reduces the problem into the numerical solution for a nonlinear partial differential equation (PDE) that contains up to the third (in 3D) power of the second-order derivatives of $\Theta$. It was later generalised by \citet{Birkin:2018nag} to more generic cases where $\delta({\bf x})$ in Eq.~\eqref{eq:jacobian} is a biased description of the true underlying matter density field. As this work does not extend the numerical algorithm to solve this PDE, we shall omit the technical details here and refer interested readers to those references. 

Once $\Theta({\bf x})$ and therefore ${\bf\Psi}({\bf x})$ are obtained, the reconstructed density field is calculated using
\begin{equation}\label{eq:delta_r}
\delta_{r} = -\nabla_{\bf q}\cdot{\bf\Psi}({\bf q}),
\end{equation}
where we have used the same symbol ${\bf\Psi}$ to denote the displacement field but note that it is now a function of the Lagrangian coordinate ${\bf q}$, and the divergence is with respect to ${\bf q}$ too. To calculate ${\bf\Psi}({\bf q})$ on a regular ${\bf q}$-grid we use the Delaunay Tessellation Field Estimator code \citep[DTFE;][]{Schaap2000,vandeWeygaert:2007ze,DTFE}, which is used to interpolate ${\bf\Psi}({\bf x})$ to a regular ${\bf q}$-grid.

\subsection{Reconstruction in redshift space}
\label{subsect:z-space-recon}

In observations, what is measured is the redshift-space coordinate, ${\bf s}$, of a particle (such as a galaxy), rather than the real-space position, ${\bf x}$. The two are related by
\begin{equation}\label{eq:s_x_relation}
    {\bf s} = {\bf x} + \frac{v_{\rm los}}{aH(a)}{\bf n},
\end{equation}
where $a$ is the scale factor, $H(a)$ is the Hubble expansion rate at $a$, ${\bf n}$ is the line-of-sight (LOS) direction and $v_{\rm los}={\bf v}\cdot{\bf n}$ is the peculiar velocity of the galaxy along the LOS direction. As a result, galaxies infalling toward massive clusters or receding from void regions can cause redshift-space distortions -- the RSD -- to the isotropic spatial distribution they would have otherwise. For it to be practically useful, therefore, the reconstruction method described above must be extended to account for the RSD effect.

We remark that Eq.~(\ref{eq:mass_conservation}) contains only ${\bf x}$ and ${\bf q}$. A similar equation that contains ${\bf s}$ and ${\bf q}$ may be obtained, allowing one to directly map between the ${\bf s}$ and ${\bf q}$ coordinates without having to worry about the ${\bf x}$ coordinate. In other words, an equation similar to Eq..~(\ref{eq:jacobian}) can be written down, and the process depicted in Sect.~\ref{subsect:r-space-recon} repeated, but with ${\bf x}$ replaced by the observed coordinate ${\bf s}$: with the assumption of no shell crossing, a unique solution of the ${\bf s}$-to-${\bf q}$ mapping is still guaranteed. However, the derivative in the left-hand side of Eq.~\eqref{eq:jacobian} is {\it formally} isotropic, whereas $\delta({\bf s})$ is anisotropic due to RSD, which means that the solutions  $\Theta({\bf s})$ and ${\bf\Psi}({\bf s})$ must be anisotropic. It is not clear whether this anisotropy would simply go away (as one would hope for) in the reconstructed density calculated using Eq.~\eqref{eq:delta_r}.

A different way to view this point is the following: our nonlinear reconstruction method starts from a set of inhomogeneously-distributed particles, and gradually moves the particles to a uniform distribution; in this process, particles can be moved in all directions as the algorithm sees necessary. If we knew how to correct ${\bf s}$ to get ${\bf x}$ exactly, the reconstruction would take two steps -- first doing that correction to get ${\bf x}$ and then solving Eqs.~(\ref{eq:jacobian}, \ref{eq:delta_r}); in the first step particles are moved along the LOS direction only, while in the second step they are moved in all directions. If we attempt to directly map ${\bf s}$ to ${\bf q}$, the first step in the above is omitted, and it is highly probable that the final solution obtained in this `crude' way differs from that of the previous, `correct', approach. One possible way to overcome this issue is to account for the additional displacements of galaxy LOS positions due to RSD by including extra terms in the equation for ${\bf\Psi}$ \citep{Nusser:1994}; the resulting equation at linear order can be solved in configuration space by finite difference \citep[e.g.,][]{Padmanabhan2012} or in Fourier space by fast Fourier transform \citep[FFT; see e.g.,][]{Burden:2014cwa}. In the latter case, the extra term breaks the translational invariance of the problem, which prevents the use of simple FFT and leads to the development of schemes to improve the solution iteratively \citep[e.g.,][]{Burden:2015pfa}. As stated, these schemes are based on the solution to a linearised equation for ${\bf\Psi}$, while we want to find a solution to the nonlinear reconstruction equation derived from Eq.~(\ref{eq:jacobian}), and it is unclear how straightforward it is to generalise them here.

An alternative method is to keep using the ${\bf x}$ coordinate in the reconstruction equation, \eqref{eq:jacobian}, but add a conversion from ${\bf s}$ to ${\bf x}$ somewhere before that equation is solved. In the Zel'dovich approximation (ZA), the displacement field ${\bf\Psi}$ and peculiar velocity field ${\bf v}$ are related as
\begin{equation}\label{eq:v_potential}
    \frac{\bf v({\bf x})}{aH} = f\nabla\Phi_v = f{\bf\Psi}\left[{\bf q}({\bf x})\right],
\end{equation}
where $\Phi_v$ is the velocity potential, $f\equiv{\rm d}\ln D_+/{\rm d}\ln a$ is the linear growth rate and $D_+$ the linear growth factor.  This suggests that, in Eq.~\eqref{eq:s_x_relation}, ${\bf s}$ can be written as a function of ${\bf x}$ and $\Theta({\bf x})$ (the latter is the potential for ${\bf\Psi}$). However, the function which connects the three quantities -- $\delta({\bf s})$, $\delta({\bf x})$ and $\Theta({\bf x})$ -- does not have an {\it a priori} known form, making it impossible to replace $\delta({\bf s})$ with $\delta({\bf x})$ in Eq.~\eqref{eq:jacobian}. This motivates a new iterative method here, which can be schematically summarised as
\begin{equation}\label{eq:iterative_method}
    {\bf x}^{(k+1)} = {\bf s} - \frac{{\bf v}^{(k)}\cdot{\bf n}}{aH(a)}{\bf n}  \longleftarrow {\bf v}^{(k)} \longleftarrow {\bf\Psi}^{(k)} \longleftarrow \delta^{(k)} \longleftarrow {\bf x}^{(k)},
\end{equation}
where $k=0,1,2,3,\cdots$ is the iteration number, and ${\bf v}^{(k)}$ the velocity field after the $k$th iteration, which is given by Eq.~(\ref{eq:v_potential}) with ${\bf\Psi}$ replaced by 
\begin{equation}\label{eq:Psi_k}
    {\bf\Psi}^{(k)} \equiv {\bf\Psi}\left({\bf x}^{(k)}\right),
\end{equation}
i.e., ${\bf\Psi}^{(k)}$ is obtained by solving Eq.~\eqref{eq:jacobian} using the particle coordinate after the $k$th iteration, ${\bf x}^{(k)}$, to calculate the density field on the right-hand side:
\begin{equation}\label{eq:delta_k}
    \delta^{(k)} \equiv \delta\left({\bf x}^{(k)}\right).
\end{equation}
At the first iteration step, $k=0$, we simply set ${\bf x}^{(0)}={\bf s}$ as our `initial guess', such that ${\bf v}^{(0)}=0$: this is equivalent to doing the reconstruction by assuming that the particles' redshift-space coordinates are identical to their real-space coordinates. Note that in Eq.~\eqref{eq:iterative_method} ${\bf s}$ is the {\it observed} coordinate in redshift space, which is {\it fixed} during the iterations.

The first equality of Eq.~(\ref{eq:iterative_method}) merits further comment as its simple form could obscure a subtle point, namely ${\bf\Psi}^{(k)}$, and hence ${\bf v}^{(k)}$, are evaluated at the ${\bf x}$, which is not known prior to the reconstruction, rather than ${\bf s}$ coordinate. This indicates that in principle this is a nonlinear equation for ${\bf x}$. Accurate solution can be found numerically once the ${\bf\Psi}^{(k)}({\bf x})$ or ${\bf v}^{(k)}({\bf x})$ fields are known. This can be done as follows: start with the approximate solution from ${\bf v}^{(k)}\approx{\bf v}^{(k)}({\bf s})$ as an initial trial solution to ${\bf x}$, ${\bf x}_{j=1}$, then take ${\bf v}^{(k)}\approx{\bf v}^{(k)}\left({\bf x}_{j=1}\right)$ to obtain an improved solution, ${\bf x}_{j=2}$, and so on, until ${\bf x}_{j}$ converges for the galaxy considered under the velocity field ${\bf v}^{(k)}$. Schematically, this can be viewed as an iterative procedure to solve the equation,
\begin{equation}\label{eq:sx-real}
    {\bf x}^{(k+1)}_{{\rm g}} = {\bf s}_{\rm g} - {\bf n}\frac{1}{aH}{\bf v}^{(k)}\left({\bf x}^{(k+1)}_{{\rm g}}\right)\cdot{\bf n},
\end{equation}
where the explicit dependence of ${\bf v}^{(k)}$ on ${\bf x}^{(k+1)}$ makes it a nonlinear algebraic equation for ${\bf x}^{(k+1)}$, and we have used the subscript `${\rm g}$' as a reminder that the coordinates are for galaxies. In practice, we used a simplified version of this scheme, described by
\begin{equation}\label{eq:sx-approx}
    {\bf x}^{(k+1)}_{\rm g} = {\bf s}_{\rm g} - {\bf n}\frac{1}{aH}{\bf v}^{(k)}\left({\bf x}^{(k)}_{\rm g}\right)\cdot{\bf n},
\end{equation}
namely the iterations for $k$ and $j$ are approximately done together. If convergence is achieved for ${\bf x}^{(k)}_{\rm g}$, we expect these two approaches to give consistent results with the latter one easier to implement. We plan to implement the full (iterative) solution to Eq.~(\ref{eq:sx-real}) as a future extension of the current pipeline.

\subsection{Simulation}
\label{subsect:simulation}
As a proof-of-concept study, in this paper we consider galaxy catalogues whose number density, $n_g$, and redshift match that of the BOSS CMASS data. More explicitly, the mock galaxy catalogues, first used in \citet{Cautun:2017tkc}, were constructed using the halo occupation distribution (HOD) model and parameters given in \citet{Manera2013}, and halo catalogues from N-body simulations of the $\Lambda$CDM model. The simulations were run using the {\sc ramses} code \citep{ramses}, employing $1024^3$ particles in a cubic box of co-moving size $1024\Mpch$, and the cosmological parameters are
\begin{equation}
    \{\Omega_m,\Omega_\Lambda,h,n_s,\sigma_8\} = \{0.281,0.719,0.697,0.971,0.8\},
\end{equation}
in which $\Omega_{m}$, $\Omega_\Lambda$ are respectively the density parameters for matter and the cosmological constant ($\Lambda$), $h\equiv H_0/(100{\rm km}{\rm s}^{-1}{\rm Mpc}^{-1})$ with $H_0$ the Hubble constant, $n_s$ is the primordial power spectrum index and $\sigma_8$ denotes the r.m.s.~matter density fluctuation smoothed on scales of $8\Mpch$. Further details of the simulations and of the HOD parameters are not very relevant for this paper, and so we opt to not report them here, but simply note that the 
galaxy catalogues correspond to redshift, $z=0.5$, have a
galaxy number density of $n_g\simeq3.2\times10^{-4}[\Mpch]^{-3}$, and that RSD effects on the coordinates of our mock galaxies were implemented by displacing the galaxies, according to their peculiar velocities from the HOD, along the three axes of the simulation box, by adopting the distant observer approximation\footnote{However, the method described here is not limited to the distant observer approximation.}: this means that for a given simulation we have produced three HOD galaxy catalogues in redshift space. 
We have five independent realisations of simulations and therefore 15 galaxy catalogues at $z=0.5$; in the analysis of Section \ref{sect:tests} we will only use the first galaxy catalogue, while all 15 are used in Section \ref{sect:application}.

\subsection{Implementation of the algorithm}
\label{subsect:implementation}

\begin{figure*}
\begin{center}
	\includegraphics[width=\textwidth]{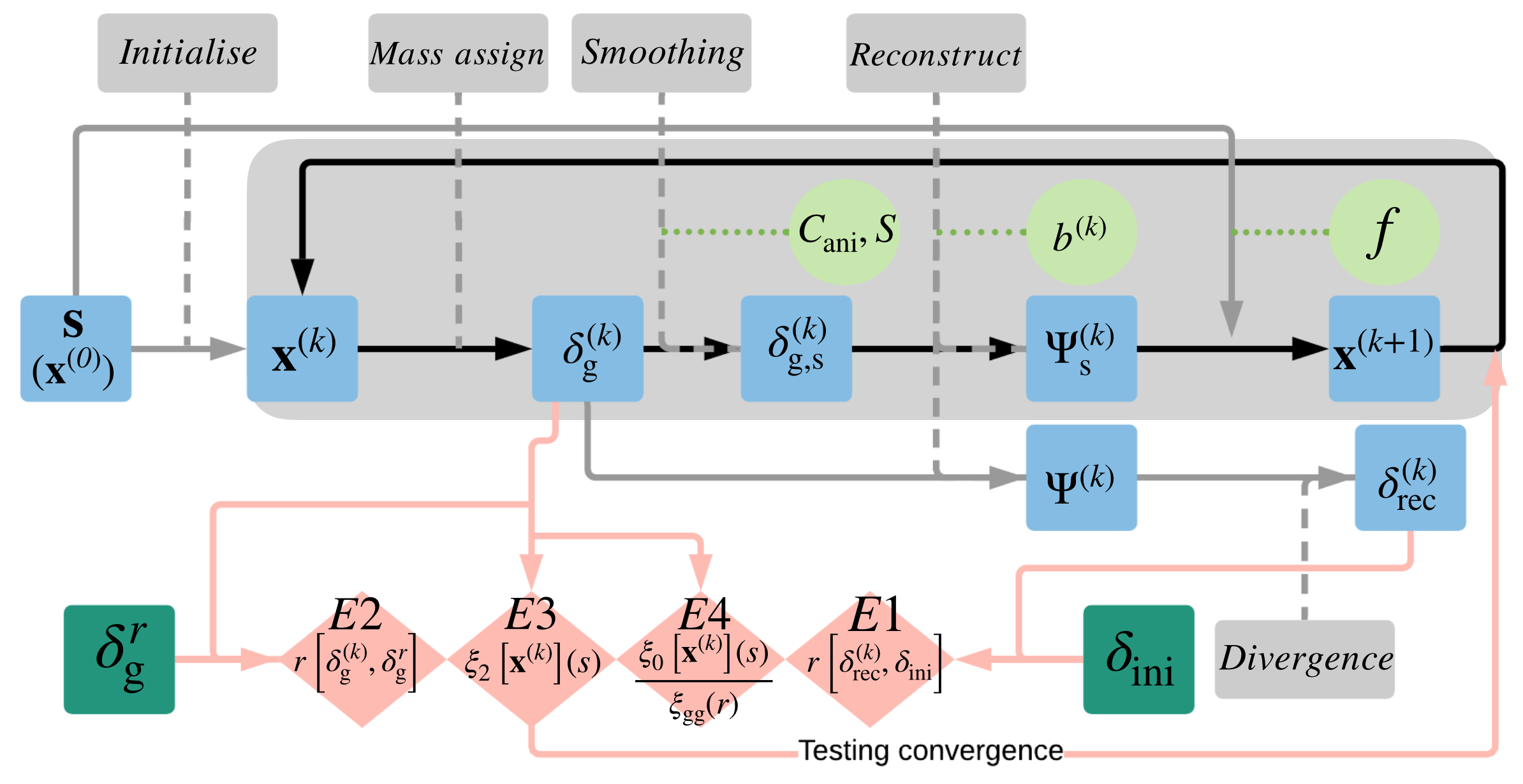}
    \caption{[{\it Colour Online}] The flowchart indicating the different steps of the iterative reconstruction pipeline introduced in this paper. The light blue boxes {(in the third and fourth rows)} are the physical quantities as input, intermediate result or output of the pipeline; the grey boxes {(in the first and last rows)} are operations that take these inputs to produce intermediate results or outputs; the pink diamonds {(in the last row)} are the estimators defined to assess the performance of reconstruction, and the dark green boxes {(in the last row)} are the real density fields which are used for evaluating two of these estimators (E1 and E2); the pink lines with arrows {(which link boxes to diamonds)} show which quantities are needed to evaluate each estimator; the light green circles indicate the parameters used in the process, which need to be tested and optimised as we will see in the next section, and the dotted green lines indicate in which operations are these parameters used. See the main text for more details.}
    \label{fig:flowchart}
\end{center}
\end{figure*}

\begin{table*}
\begin{tabular}{@{}|l|l|l|}
\hline\hline
Symbol & Physical meaning & Value \\
\hline
${\bf x}$ & real-space galaxy coordinate & $-$ \\
$r$ & real-space distance & $-$ \\
${\bf s}$ & redshift-space galaxy coordinate & $-$ \\
$s$ & redshift-space distance & $-$ \\
${\bf q}$ & initial (Lagrangian) coordinate & $-$ \\
${\bf x}^{(k)}$ & reconstructed real-space galaxy coordinate ($k$th iteration) & $-$ \\
${\bf\Psi}^{(k)}_{\rm S}$ & displacement field from reconstruction on smoothed galaxy density field ($k$th iteration) & $-$ \\
${\bf\Psi}^{(k)}$ & displacement field from reconstruction on un-smoothed galaxy density field ($k$th iteration) & $-$ \\
$\delta_{\rm ini}$  & initial matter density field & $-$ \\
$\delta^r_{\rm g}$ & final real-space galaxy density field & $-$ \\
$\delta^s_{\rm g}$ & final redshift-space galaxy density field & $-$ \\
$\delta_{\rm rec}$ & reconstructed matter density field from final real-space galaxy catalogue & $-$ \\
$\delta_{\rm rec}^{(k)}$ & reconstructed matter density field from reconstructed real-space galaxy catalogue ($k$th iteration) & $-$ \\
$\delta_{\rm g}^{(k)}$ & galaxy density field of reconstructed real-space galaxy catalogue ($k$th iteration) & $-$ \\
$\delta_{\rm g,S}^{(k)}$ & smoothed galaxy density field of reconstructed real-space galaxy catalogue ($k$th iteration) & $-$ \\
$r[a,b]$ & cross correlation coefficients between fields $a$ and $b$ & $-$ \\
$\xi_{\rm gg}(r)$ & real-space galaxy auto-correlation function & $-$ \\
$\xi_{\rm gm}(r)$ & real-space galaxy-matter cross correlation function & $-$ \\
$\xi^s_{\rm gg}({\bf s})$ & redshift-space galaxy auto-correlation function & $-$ \\
$\xi_{0,2,4}(s)$ & redshift-space galaxy correlation function monopole, quadrupole and hexadecapole & $-$ \\
$K$ & value of iteration number $k$ at convergence & $3$-$6$ \\
\hline
$f$ & linear growth rate & $0.735$ \\
$b^{(k)}$ & linear galaxy bias ($k$th iteration) & - 
\\
$b_{\rm sim}$ & linear galaxy bias measured in simulation & $1.95$ \\
\hline
${n}_{\rm g}$ & galaxy number density & $3.2\times10^{-4}\left[\Mpch\right]^{-3}$ \\
${\rm d}x$ & reconstruction grid cell size & $2\Mpch$ \\
$S$ & isotropic Gaussian smoothing scale & $9\Mpch$ \\
$C_{\rm ani}$ & anisotropic smoothing parameter & $1.0$ \\
\hline
E1 & $r\left[\delta^{(k)}_{\rm rec},\delta_{\rm ini}\right]$ & $-$ \\
E2 & $r\left[\delta^{(k)}_{\rm g},\delta_{\rm g}^r\right]$ & $-$ \\
E3 & $\xi_2\left[{\bf x}^{(k)}\right](s)$ & $-$ \\
E4 & $\xi_0\left[{\bf x}^{(k)}\right](s)/\xi_{\rm gg}(r)$ & $-$ \\
$R(s)$ & $\xi_0(s)/\xi_0\left[{\bf x}^{(k)}\right](s)$ & $-$ \\
$R'(s)$ & $\xi_0(s)/\xi_{\rm gg}(r)$ & $-$ \\
\hline\hline
\end{tabular}
\caption{A short summary of the symbols used throughout this paper. The first block (from ${\bf x}$ to $K$) contains the various quantities used in the reconstruction process, the second block ($f$ to $b_{\rm sim}$) are physical parameters related to the galaxy catalogues, the third block ($n_{\rm g}$ to $C_{\rm ani}$) are technical parameters used in the reconstruction, and the last block (E1 to $R'(s)$) are estimators defined to check the convergence of reconstruction. The first column contains the symbols, the second column their physical meaning and the last column the default values (a `$-$' is used for quantities without default values). We find that for estimators E1 and E2 the number of iterations required before convergence is generally smaller than for estimators E3 and E4, and so a range of values is given for $K$.}
\label{table:params}
\end{table*}

The description of our iterative reconstruction algorithm in the previous subsection is quite schematic, and therefore in this subsection we give more technical details of its implementation. The presentation here shall follow the logic as depicted in the flowchart, Fig.~\ref{fig:flowchart}, and for clarity we also list all the physical or numerical parameters, and their meanings, in Table \ref{table:params}.

The main ingredients of the reconstruction algorithm are listed below (where a superscript $^{(k)}$ denotes the corresponding quantities after the $k$th reconstruction iteration):
\begin{enumerate}
\item {\it Creating the galaxy density field $\delta_{\rm g}^{(k)}$ on a uniform grid using the approximate real-space coordinates of the galaxies, ${\bf x}^{(k)}$}. 
This is done using the triangular-shaped cloud (TSC) mass assignment scheme implemented in the DTFE public code \citep{DTFE}. Note that we do {\it not} use actual Delaunay tessellation to calculate the density field, as it has been shown by \citet{Birkin:2018nag} -- and checked again in this project -- that this leads to a poorer reconstruction performance. 

The size of the uniform grid on which $\delta_{\rm g}^{(k)}$ is calculated has some effect on the reconstruction result, and in this work we have adopted a grid with $512^3$ cells, i.e., with cell size ${\rm d}x=2\Mpch$, because using a grid with even higher resolution does not make a significant difference \citep{Birkin:2018nag}.

\item {\it Calculating the displacement field ${\bf\Psi}$ and performing reconstruction}. Here things become a bit tricky: even though we are trying to simultaneously do reconstructions of the initial density field and the real-space galaxy coordinates, the optimal technical specifications are not the same in the two cases. As a result, we actually do two reconstruction calculations of ${\bf\Psi}$ for a given $\delta_{\rm g}^{(k)}$ field, both using the {\sc ecosmog} code developed by \citet{Shi:2017gqs} and \citet{Birkin:2018nag}.

In the first calculation, the objective is to undo the RSD and thus to bring the galaxy coordinates, ${\bf x}^{(k)}$, closer to their true real space values, ${\bf x}$. Here, our concern is that the stretching effects of FoG could lead to erroneous estimation of the large-scale density field, causing worse performance of the method. To reduce its impact, we follow \citet{Hada:2018fde} and calculate the density field, $\delta_{\rm g}^{(k)}$, using an anisotropic smoothing function. The filtering function is chosen to be a skewed Gaussian that has a different smoothing length along the line-of-sight direction, and the smoothed galaxy density field is given, in Fourier space, as\footnote{Note the slight abuse of notation here: $k$ is used both to denote the iteration number and to represent the wave number/vector in Fourier space.}
\begin{equation}
		\tilde{\delta}^{(k)}_{\rm g,S}(k) = 
		\tilde{\delta}^{(k)}_{\rm g} \tilde G({\bf k}) \equiv 
		\tilde{\delta}^{(k)}_{\rm g} \exp\left[-\left(k_{n}^2 S_{n}^2 + k_{p}^2S_{p}^2\right)\right],
		 \end{equation}
where ${\bf k}$ is the wave number with $k_n$ and $k_{p}$ representing the wave numbers along the line-of-sight and perpendicular to it. The functions $\tilde{\delta}^{(k)}_{\rm g}$, $\tilde G({\bf k})$ are the Fourier transformations of $\delta^{(k)}_{\rm g}$ and the filter mentioned above. This introduces two extra parameters for the algorithm, $S_n$ and $S_p$, and in what follows we express them by $S=S_p$ (the smoothing length perpendicular to LOS) and a dimensionless parameter $C_{\rm ani}\equiv S_{n}/S$, 
{with $C_{\rm ani}>1$ representing a larger smoothing length along the LOS}. The calculation from here on is similar as before, but with $\delta_{\rm g,S}^{(k)}$ instead of $\delta_{\rm g}^{(k)}$ being fed into {\sc ecosmog}, and $b^{(k)}$ is applied again to convert this to an approximated nonlinear matter density field\footnote{In principle the $b^{(k)}$ parameters used here can be different from the ones used in the first calculation above, but in our implementation we have used the same $b^{(k)}$ for a given $k$ iteration.}. The displacement field obtained here is denoted as ${\bf\Psi}^{(k)}_{\rm S}$, from which we can derive the `improved' real-space galaxy coordinates, ${\bf x}^{(k+1)}$, as
\begin{equation}
    {\bf x}^{(k+1)} = {\bf s} - f{\bf\Psi}^{(k)}_{\rm S},
\end{equation}
where $f$ is the linear growth rate introduced above, which we take as a scale-independent (but time-dependent) constant, as is the case for $\Lambda$CDM and several dark energy and modified gravity models. At $z=0.5$, the equation
\begin{equation}
	f(z) \simeq \left[\Omega_{m}(z)\right]^{0.55}
\end{equation}
is a very good approximation, which gives a value of $f=0.735$, in good agreement with numerical result obtained by using the cosmological parameters given above. However, in the actual calculation we have left $f$ to be a free parameter to be varied because its value is {\it a priori} unknown in observations.

In the second calculation, the aim is to obtain the reconstructed matter density field, $\delta_{\rm rec}^{(k)}$, using the relation
\begin{equation}
    \delta_{\rm rec}^{(k)} = -\nabla_{\bf q}\cdot{\bf\Psi^{(k)}},
\end{equation}
where the displacement field at the $k$th iteration, ${\bf\Psi}^{{(k)}}$, is calculated by applying {\sc ecosmog} to $\delta_{\rm g}^{(k)}/b^{(k)}$, without doing any smoothing (which would degrade the performance; see below and \citealt{Birkin:2018nag}). Here $b^{(k)}$ is the linear bias parameter such that $\delta_{\rm g}^{(k)}/b^{(k)}$ is an approximation to the nonlinear matter density field; note that here we assume different values of $b^{(k)}$ need to be used in the different iterations. 

\item {\it Checking for convergence}. As an iterative solution scheme, we need a criterion (or a set of criteria) to decide when the iterations can be stopped. Usually, convergence is deemed to be achieved if the error (defined in whatever way) is reduced to below some preset tolerance, e.g., some small number. The problem at hand is more complicated in that, {\it a priori}, there is no `target' solution to be used to clearly define the `error'. Therefore, here we opt for a set of loose criteria for convergence:

C1: a set of estimators obtained from the reconstruction outcome `stabilise' and do not change further with increasing number of iterations ($k$). This is a generic convergence criterion which is essential for the method to work, and we require it to be satisfied for any estimator to be considered. This criterion is also practically useful, as it applies to both statistics extracted directly from observations (such as estimator E3 to be introduced below) and theoretical quantities that are only known in controlled experiments, such as simulations. The latter, however, are also helpful since they offer other ways to assess the performance of and to determine the optimal parameters for the reconstruction; for this reason, we also introduce two more convergence criteria that apply only to theoretical quantities:

C2: assuming that convergence is achieved after iteration $k=K$, then reconstructed matter density field $\delta_{\rm rec}^{(K)}$ is `closer' to the initial density field $\delta_{\rm ini}$ than any of the pre-convergence results, $\delta_{\rm rec}^{(k)}$, $\forall k<K$; here $\delta_{\rm ini}$ is a theoretical quantity;

C3: the reconstructed galaxy coordinates ${\bf x}^{(K)}$ are `closer' to the true real-space galaxy coordinates ${\bf x}$ than any pre-convergence results, ${\bf x}^{(k)}$, $\forall k<K$; here ${\bf x}$ is a theoretical quantity.

It is not our objective to be very quantitative in defining convergence, and instead we simply check that `by eye', i.e., we stop the iterations if the statistic or estimator of interest has stabilised and does not change significantly after further iterations. Four estimators are defined, which can be constructed from the reconstruction outcome, to allow us to test these criteria. Different estimators may need different numbers of iterations before convergence, and these are shown in Table \ref{table:params}.

For Criterion C2, we use the usual cross correlation coefficient, $r$, between the reconstructed and initial density fields, to characterise the similarity between them. The correlation coefficient between any two fields $\delta_a$, $\delta_b$ is defined as 
\begin{equation}
r[a,b] \equiv \frac{\tilde{\delta}_a\tilde{\delta}_b^{\ast} + \tilde{\delta}_a^{\ast}\tilde{\delta}_b}{2\sqrt{\tilde{\delta}_a\tilde{\delta}_a^{\ast}}\sqrt{\tilde{\delta}_b\tilde{\delta}_b^{\ast}}},
\end{equation}
where $\tilde{\delta}_a$, $\tilde{\delta}_b$ are the Fourier transforms of $\delta_a$ and $\delta_b$ and a superscript $^*$ denotes \textcolor{black}{taking the complex conjugate.}
A value of $r[a,b]=1$ means perfect correlation while $r[a,b]=0$ means that $a$ and $b$ are completely random. In other words, for C2 we would like that $r\left[\delta^{(K)}_{\rm rec},\delta_{\rm ini}\right]$ to be closer to 1 than $r\left[\delta^{(k)}_{\rm rec},\delta_{\rm ini}\right]$, for $\forall k<K$. Since $r[a,b]$ is a function of scale, or Fourier wavenumber, $k$, ideally we hope the above applies for all wavenumber values or, if that is not possible, at least for the range of wavenumbers of most interest to us.

For C3 we have defined a similar estimator by cross-correlating $\delta_{\rm g}^{(k)}$ with the final real-space galaxy density field, $\delta_{\rm g}^r$, and requiring that $r\left[\delta^{(K)}_{\rm g},\delta_{\rm g}^r\right]$ is closer to $1$ than $r\left[\delta^{(k)}_{\rm g},\delta_{\rm g}^r\right]$, $\forall k<K$.

We have also defined two more estimators based on the argument that, if ${\bf x}^{(K)}$ is close enough to ${\bf x}$, then the two-point correlation functions obtained from these two galaxy catalogues should also be close to each other. In particular, the RSD-induced anisotropy in the two-point correlation function of the redshift-space (${\bf s}$) galaxy catalogue should be largely removed in the reconstructed, ${\bf x}^{(K)}$, galaxy catalogue. Therefore, we require that $\xi_2\left[{\bf x}^{(K)}\right]$, the quadrupole of the two-point galaxy correlation function of the ${\bf x}^{(K)}$ catalogue, be closer to $0$ than $\xi_2\left[{\bf x}^{(k)}\right]$, $\forall k<K$\footnote{Note that we have used [] to highlight that ${\bf x}^{(k)}$ is not an argument of $\xi_2$ but simply is a symbol to represent a given galaxy catalogue. The proper argument for $\xi_2(s)$, not shown here to lighten the notation, is the galaxy pair distance in redshift space, $s$. As above, ideally we would like $\xi_2\left[{\bf x}^{(K)}\right]$ to be close to $0$ on all scales or, if it is not possible, at least in the scales of most interest to us.}.

In addition, we would also expect that $\xi_0\left[{\bf x}^{(K)}\right]$, the monopole of the two-point galaxy correlation function of the ${\bf x}^{(K)}$ catalogue, to be close to the real space galaxy correlation function $\xi_{\rm gg}$. Therefore, a further requirement is that the ratio $\xi_0\left[{\bf x}^{(K)}\right]/\xi_{\rm  gg}$ be closer to 1 than $\xi_0\left[{\bf x}^{(k)}\right]/\xi_{\rm gg}$, $\forall k<K$. In this paper, we measure $\xi_0$ and $\xi_2$ using the publicly available code `Correlation Utilities and Two-point Estimators' \citep[{\sc cute;}][]{Alonso2012}.

Note that in certain situations we may need to loosen the above requirements. Taking $\xi_2\left[{\bf x}^{(k)}\right]$ for example, it is possible that for some intermediate $k<K$ the result coincidentally gets very close to zero (this may happen if $\xi_2\left[{\bf x}^{(k)}\right]$ oscillates around $0$ for increasing $k$). Therefore it is always safe to try a couple more iterations even if the result seems to have converged.

\item {\it Finalising the code}. Finally, once convergence is deemed to have been achieved, we stop the iterations at $k=K$. 

In what follows, to avoid carrying cumbersome notations everywhere, we shall call the four estimators introduced above E1, E2, E3 and E4, respectively. Note that out of these estimators, only E3 is applicable in real observations because the other three all require something that only exists in simulations in their definitions -- $\delta_{\rm ini}$ for E1, $\delta_{\rm g}^r$ for E2 and $\xi_{\rm gg}(r)$ for E4. As a result, the latter estimators are mainly used in this work as theoretical tools to demonstrate the performance of the iterative reconstruction algorithm, and to determine the optimal technical parameters.

On the other hand, E3 can be estimated using observational data alone. Therefore, our objective in the following parts of this paper is to check what is the potential of using E3 alone to determine the `best-fit' values of the physical parameters, such as $f$ and $b^{(k)}$, and to do the RSD reconstruction. If $f, b$ could be precisely determined in this process, then that would be an additional benefit of this new algorithm, along with simultaneously giving us approximate reconstructions of the initial (linear) and final (nonlinear) matter density fields and the final real-space galaxy density field (or coordinates). These will turn out to be very useful information as we exemplify and discuss later. In the less ideal scenario, if $f,b$ could not be accurately determined (for example because the reconstruction outcome is not very sensitive to them), then the other benefits would remain.

Note that we can also use higher-order multipole moments, such as the hexadecapole $\xi_{4}\left[{\bf x}^{(k)}\right](s)$, as more estimators to check the convergence, namely $\xi_{4}\left[{\bf x}^{(K)}\right]$ must be closer to $0$ than $\xi_{4}\left[{\bf x}^{(k)}\right]$, $\forall k<K$. These have the advantage that they can be obtained from real observational data. In particular, it would be interesting to see if they offer consistent (or complementary) constraints on the physical parameters, such as $f$ and $b$. However, for our galaxy catalogues the number density $n_{\rm g}\simeq3.2\times10^{-4}[\Mpch]^{-3}$ is too low and the measurements of $\xi_{4}\left[{\bf x}^{(k)}\right]$ are too noisy. Therefore, we shall leave a check of the impacts of such additional estimators to a future work, where we'll test the reconstruction algorithm using 
galaxy catalogues with various number density cuts.

\end{enumerate}

\section{Reconstruction tests and performance}
\label{sect:tests}

We tested the reconstruction pipeline for a large number of combinations of the physical and technical parameters, $\{f, b^{(k)}, S, C_{\rm ani}\}$, in which $b^{(k)}$ were allowed to vary with the iteration number, $k$, in order to settle to the most optimal choices of $S, C_{\rm ani}$ and to explore the potential of constraining $f,b$ as a byproduct of reconstruction. The optimal values for these parameters are summarised in the last column of Table \ref{table:params}, and in this section we will show the impacts of varying these parameters on the reconstruction performance. As we have a relatively large parameter space, we shall only vary a subset of them -- while fixing the others to the optimal values -- at a given time.

\begin{figure}
    \hspace*{-0.72cm}
	\includegraphics[width=1.17\columnwidth]{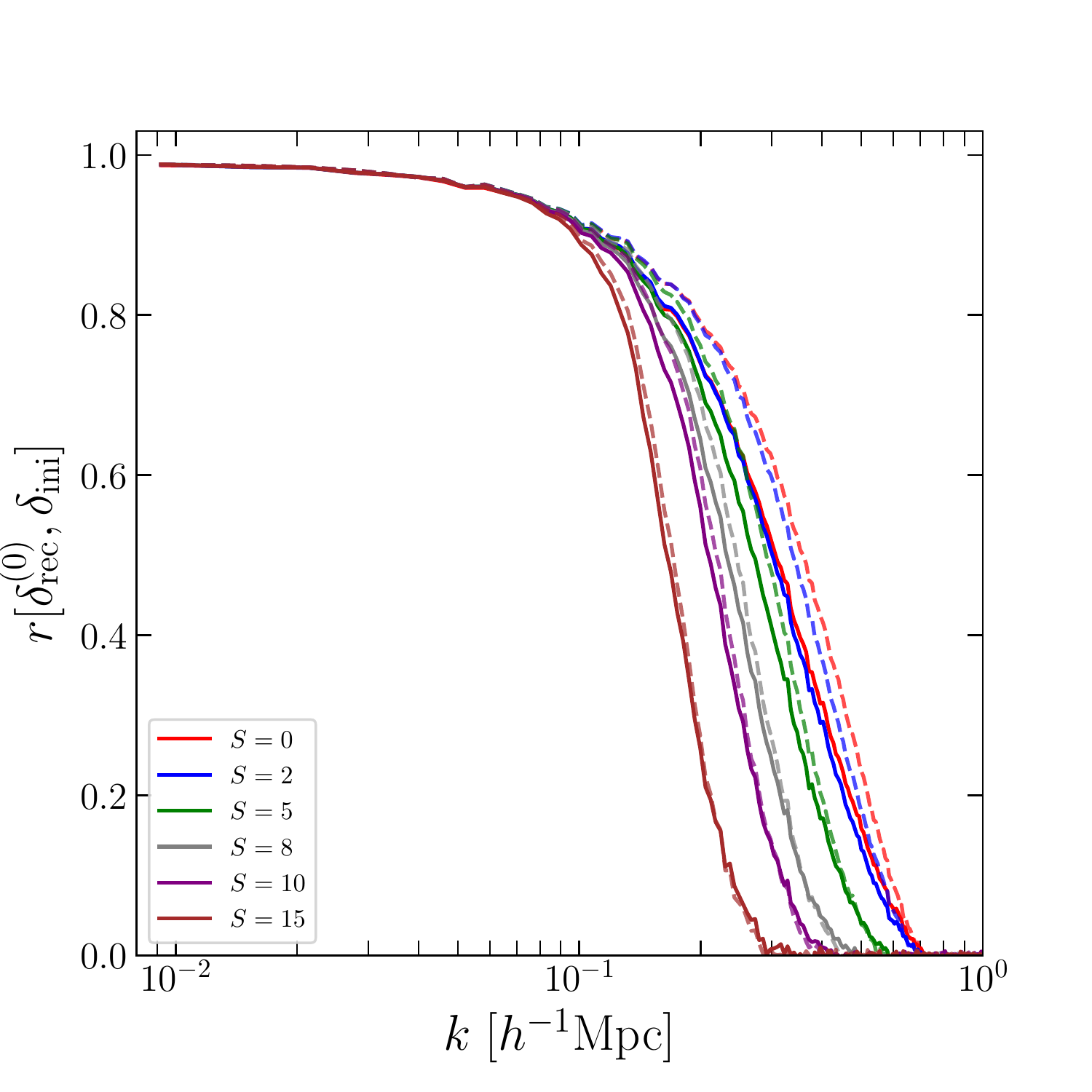}
	\vskip -.3cm
    \caption{{[{\it Colour Online}]} The 
    cross correlation coefficients of the initial density field with the reconstructed matter density field from the real-space galaxy catalogue ($r\left[\delta_{\rm rec},\delta_{\rm ini}\right]$; dashed lines) and with the reconstructed matter density field from the redshift-space galaxy catalogue ($r\left[\delta_{\rm rec}^{(0)},\delta_{\rm ini}\right]$; solid lines) using {\it no} iterations. The
    various coloured lines correspond to
    the results for which the galaxy density field has been smoothed by a skewed Gaussian filter with $C_{\rm ani}=1.0$ and $S=0$ (no smoothing; red; {rightmost curve}), $2$, $5$, $10$ and $15\Mpch$ (brown; {leftmost curve}). The bias parameter used here is $b^{(0)}=2.0$.
    }
    \label{fig:Real_red_Comparison}
\end{figure}

Before going to the details, in Fig.~\ref{fig:Real_red_Comparison} we present a quick visual inspection of the impact of RSD on the  reconstruction performance. The red dashed line is estimator E1, $r\left[\delta_{\rm rec},\delta_{\rm ini}\right]$, between the initial matter density field, $\delta_{\rm ini}$, and the reconstructed matter density field, $\delta_{\rm rec}$, from the final galaxy catalogue in real space. The red solid line differs by replacing $\delta_{\rm rec}$ with $\delta_{\rm rec}^{(0)}$, which is the reconstructed matter density field from the zeroth-iteration of our RSD reconstruction, namely by incorrectly assuming that the redshift-space coordinates of the galaxies are also their real-space coordinates without any corrections, or equivalently applying the reconstruction code of \citet{Birkin:2018nag} directly to our redshift-space galaxy catalogue without using iterations. We can see that not cleaning up RSD effects causes the correlation to become smaller than in real-space reconstruction, which is as expected. However, the impact is mild, which is perhaps because of the relatively low galaxy number density used here. As a result, we expect that any improvement by iterative reconstruction will be mild as well (but note that both conclusions might not hold for galaxy catalogues with much higher $n_{\rm g}$.)

The dashed and solid lines with other colours in Fig.~\ref{fig:Real_red_Comparison} are very similar, but they correspond to results where both the real- and the redshift-space galaxy density fields are further smoothed -- after the TSC mass assignment -- using the skewed Gaussian filter described above, with $C_{\rm ani}=1$, $S=2$ (blue), $5$ (green), $8$ (grey), $10$ (purple) and $15\Mpch$ (brown). Notice that the red lines described above are results from unsmoothed galaxy density field and correspond to $S=0$. We can see a clear trend that smoothing the galaxy density field leads to poorer outcomes of the reconstruction (as mentioned earlier), which is because the smoothing effectively suppresses the small-scale features of the density field. This is why when describing the flowchart (Fig.~\ref{fig:flowchart}) above we emphasised that smoothing is  used in calculating the displacement field ${\bf\Psi}_{\rm S}^{(k)}$ which is needed to correct galaxy coordinates, and not in calculating the displacement field ${\bf\Psi}^{(k)}$ which is used to obtain the reconstructed matter density field. Also note that for all tests in Fig.~\ref{fig:Real_red_Comparison} we have used $b^{(0)}=2.0$ and that $f$ is not used here.

Another interesting feature in Fig.~\ref{fig:Real_red_Comparison} is that, as the smoothing length $S$ increases, the difference between real and redshift-space reconstructions reduces, and with $S=15\Mpch$ (brown lines) the two cases almost agree perfectly with each other. This is again not surprising given that the effect of RSD is to shift galaxy positions while smoothing to certain extent undoes that shift. However, this is at a price of suppressing small-scale features and leading to poorer reconstruction results for both real and redshift spaces.

We now present the results of a wide range of tests, to illustrate the (lack of) impacts of varying different physical and technical parameters used in the iterative procedure on the estimators defined in the previous section. As mentioned above, these parameters serve both as fitting parameters used to identify the optimal reconstruction specifications, as well as informative vehicles that can provide valuable insights into the formation of large-scale structures.

\subsection{Smoothing parameters $S$ and $C_{\rm ani}$}
\label{subsubsect:vary_SC}

\begin{figure*}
\centering
	\includegraphics[width=1\textwidth]{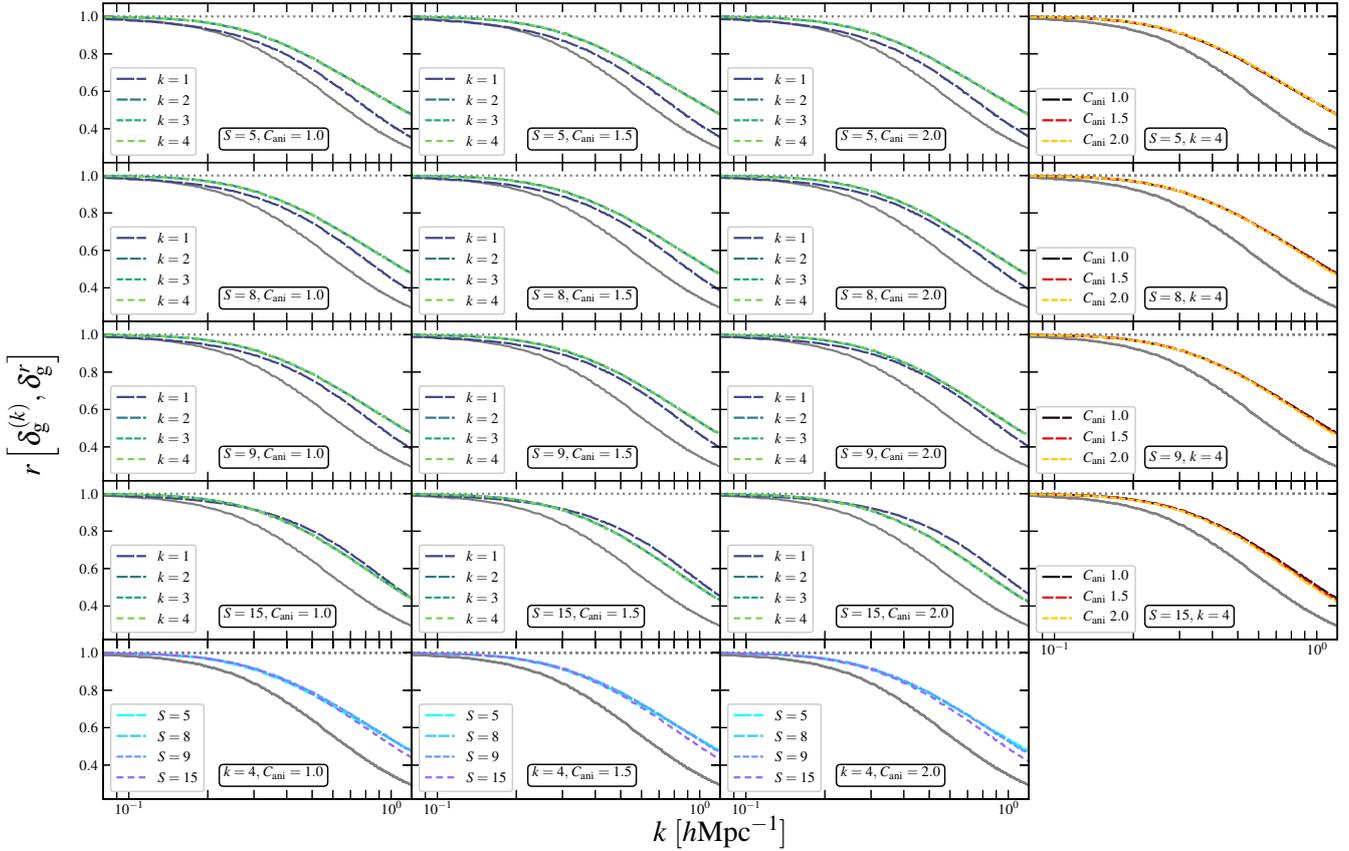}
	\vskip -.3cm
    \caption{{[{\it Colour Online}]} Estimator E2, $r\left[\delta_{\rm g}^{(k)}, \delta_{\rm g}^r\right](k)$ for various combinations of technical parameters $S$ and $C_{\rm ani}$. Each of the first four rows includes tests using a fixed $S$, which takes value of $5, 8, 9$ and $15\Mpch$ respectively; each of the first three columns corresponds to tests using a fixed $C_{\rm ani}$, which takes value of $1.0, 1.5$ and $2.0$ respectively. The $4\times3$ block of subpanels on the top left show how E2 changes with increasing number of iterations for a given $(S,C_{\rm ani})$. Each of the three subpanels at the bottom compares the results for fixed $C_{\rm ani}$ and varying $S$, at the last iteration; each of the four subpanels on the far right compares the results for fixed $S$ and varying $C_{\rm ani}$, again at the {\it last} iteration. {The sparseness of the dahses lines increases with $k$, $S$ or $C_{\rm ani}$ in the three different regions (see the legends).} The {grey solid} lines are the same in all subpanels and show $r\left[\delta^s_{\rm g},\delta^r_{\rm g}\right](k)$, which is the cross correlation between the final galaxy density fields in redshift and real spaces. {The grey dotted lines are zero.}}
    \label{fig:Cor_real_SC_sum}
\end{figure*}

\begin{figure*}
\centering
	\includegraphics[width=1\textwidth]{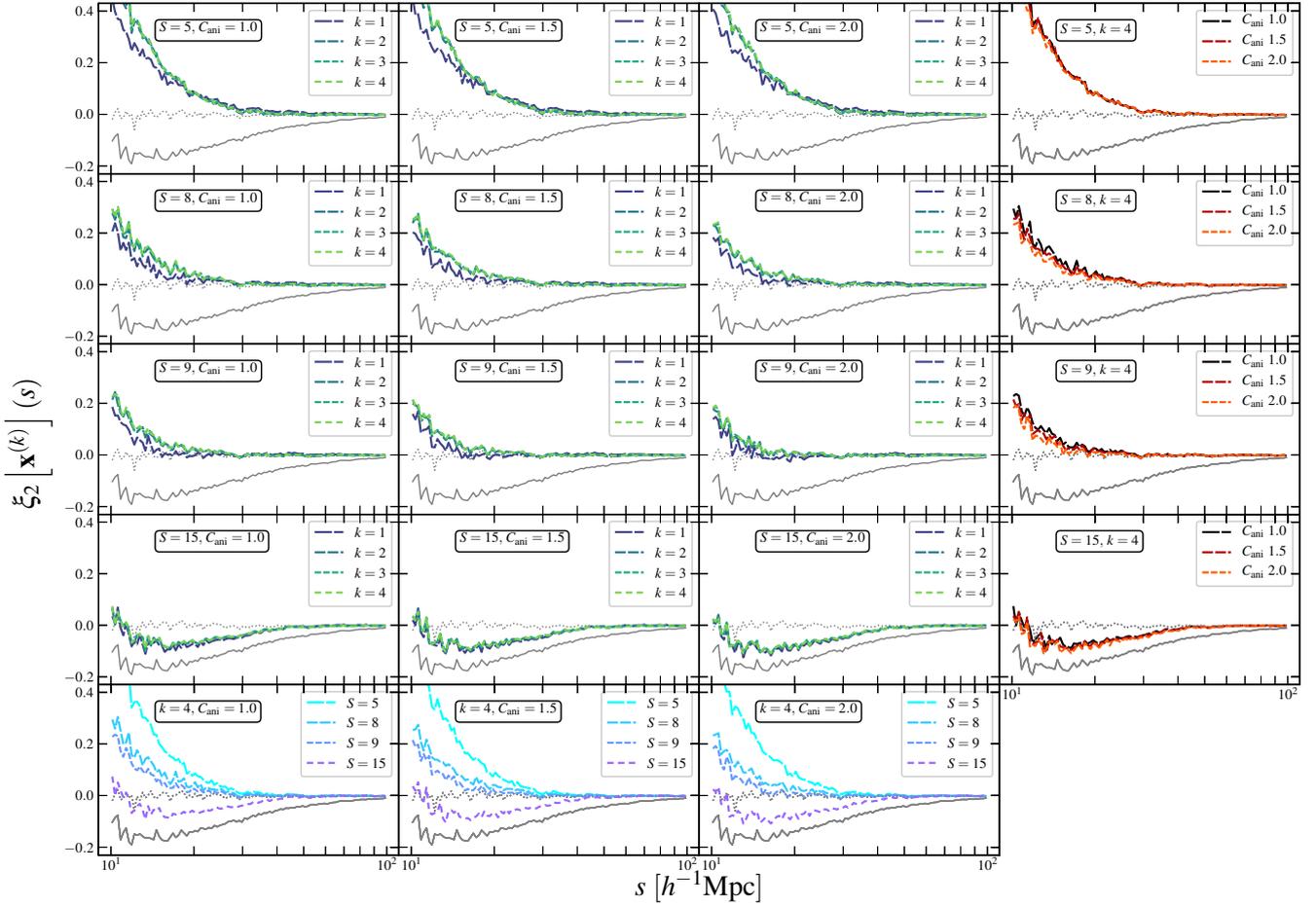}
	\vskip -.3cm
    \caption{{[{\it Colour Online}]} The same as Fig.~\ref{fig:Cor_real_SC_sum}, but for estimator E3, $\xi_2\left[{\bf x}^{(k)}\right](s)$. The {grey dotted} and {solid} lines are the same in all subpanels: the {former}, which is very close to $0$ on the entire range of scales, is the quadrupole moment measured from the real-space galaxy catalogue at $z=0.5$, while the {latter}, which is negative in the whole $s$ range, is that measured from the redshift-space galaxy catalogue at $z=0.5$.}
    \label{fig:quadrupole_SC_sum}
\end{figure*}

\begin{figure*}
\centering
	\includegraphics[width=1\textwidth]{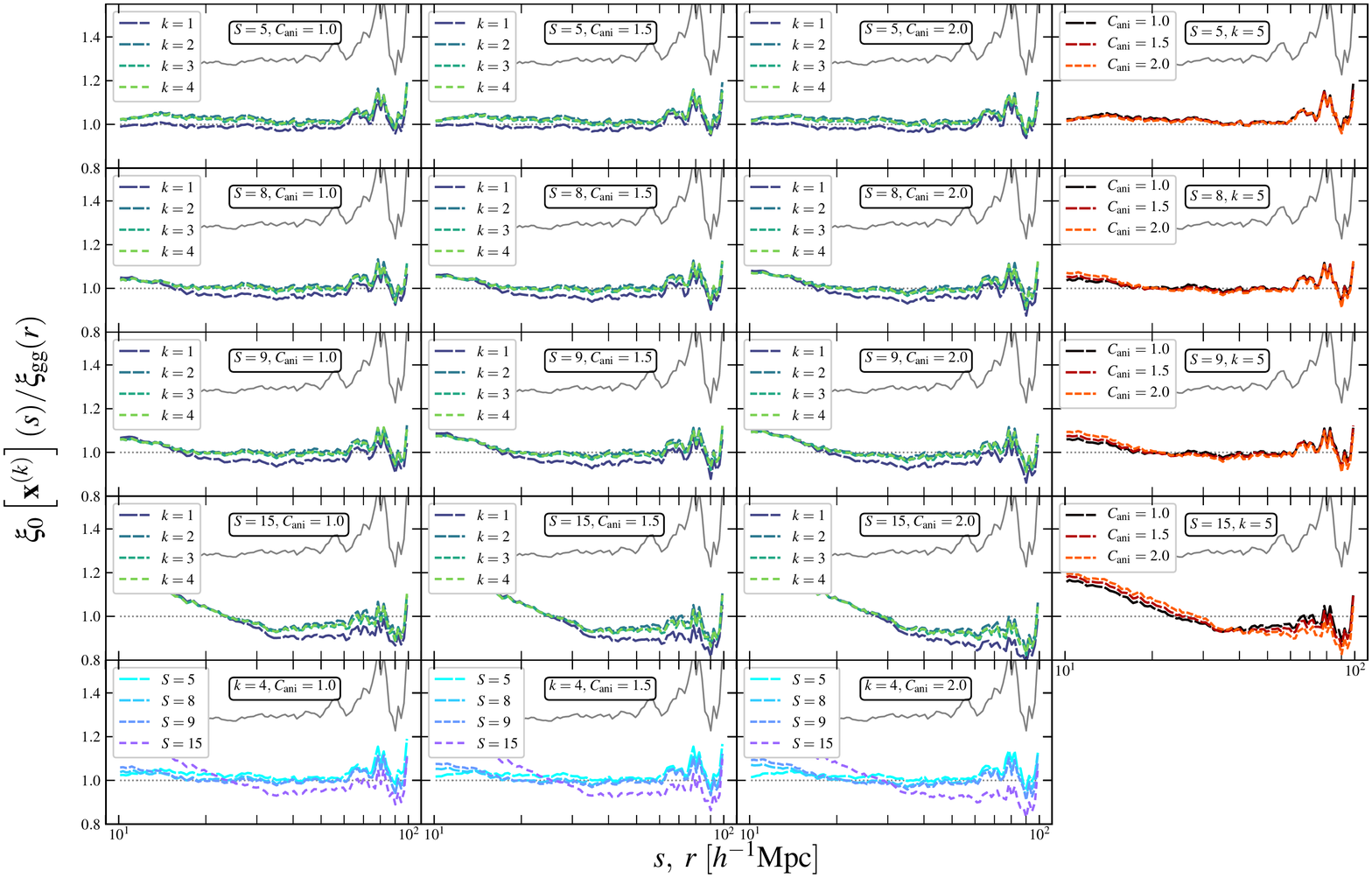}
	\vskip -.3cm
    \caption{{[{\it Colour Online}]} The same as Fig.~\ref{fig:Cor_real_SC_sum}, but for estimator E4, $\xi_0\left[{\bf x}^{(k)}\right](s)/\xi_{\rm gg}(r)$. The straight {dotted} and the wiggly {solid} grey lines are the same in all subpanels: the former is the constant $1.0$ to guide the eyes, while the latter is the ratio between the monopole moments measured from the redshift- and real-space galaxy catalogues at $z=0.5$. 
    }
    \label{fig:monopole_SC_sum}
\end{figure*}

Let us first discuss the the smoothing-related parameters $S$ and $C_{\rm ani}$. $S$ represents the overall smoothing length, while $C_{\rm ani}$ characterises the amount of anisotropic smoothing, with $C_{\rm ani}=1.0$ indicating no anisotropy in the smoothing function and $C_{\rm ani}>1.0$ indicating a longer smoothing length along the line-of-sight direction to suppress the impact of the FoG effect.

In Fig.~\ref{fig:Cor_real_SC_sum} we compare estimator E2 constructed from the reconstruction outputs for $4\times3$ combinations of $(S,C_{\rm ani})$: four choices of $S$ -- $5, 8, 9$ and $15\Mpch$ -- and three choices of $C_{\rm ani}$ -- $1.0, 1.5$ and $2.0$. The results for each of the 12 combinations are shown in one of the 12 subpanels on the top left of Fig.~\ref{fig:Cor_real_SC_sum}, and in each subpanel the different lines are the results after different numbers of iterations, $k$, $r\left[\delta^{(k)}_{\rm g},\delta^r_{\rm g}\right]$, with $k=1,2,3,4$. As a comparison, the dashed line represents $r\left[\delta^s_{\rm g},\delta^r_{\rm g}\right]$, namely the cross correlation between the final real and redshift-space galaxy density fields. Within each row, the smoothing scale perpendicular to the LOS, $S_p$, is fixed while $S_n$, as being the product of $S$ and $C_{\rm ani}$, changes across the columns.

By comparing the different columns in a given row in Fig.~\ref{fig:Cor_real_SC_sum}, it is evident that the effect of $C_{\rm ani}$ on estimator E2 is only significant for first few iterations. For $k = 4$, the difference between $C_{\rm ani}=1.0, 1.5$ and $2.0$ is much smaller. The convergence criterion C1 is satisfied by all tests, regardless of the value of $C_{\rm ani}$. 

The overall behaviour of $r\left[\delta^{(k)}_g,\delta^r_{\rm g}\right]$ for small \revise{smoothing scale} is as expected. One can consider the FoG effect as some `redistribution' of galaxies around the centres of their host haloes, where virial motions of the former can lead to the measured galaxy coordinates differing from their actual values by an amount much larger the radii of the dark matter haloes. If uncorrected, this could cause a galaxy 1 which is closer to us than another galaxy 2 in real space to actually appear to be farther away than galaxy 2 in redshfit space. In other words, a `shell crossing' takes place due purely to the use of redshift space coordinates, and this violates one of the basic assumptions of the reconstruction method, which leads to a degraded performance of the latter. This impact can be alleviated if the galaxy density field is smoothed using a large filter, whose size is at least comparable to the typical peculiar-velocity-induced changes of galaxy distances in redshift space. \revise{Choosing a too large smoothing scale reduces the correlation due to the loss of information, so an optimal scale must be found that balances the FoG effect without compromising the accuracy of the reconstruction.}

The physical reasoning given in the above paragraph is supported by the following observation of Fig.~\ref{fig:Cor_real_SC_sum}, namely in the cases of \revise{smaller smoothing lengths, for both $S$ and $C_{\rm ani}$, the cross correlation $r\left[\delta^{(k)}_g,\delta^r_{\rm g}\right]$ is generally larger.} Since the smoothing length $S_n$ along the LOS direction is the product of $S$ and $C_{\rm ani}$, we have \revise{$S_n=9$ for $\left(S, C_{\rm ani}\right)=(9, 1)$} which are sufficiently large to smooth out the FoG effect (notice that for typical galaxies the LOS velocities are smaller than $2000$ km/s so that ${\bf v}\cdot{\bf n}/aH\lesssim20\Mpch$). \revise{Overall, we find that for a fixed $C_{\rm ani}$ value,} increasing $S$ (or equivalently $S_p$) leads to \revise{a slightly faster convergence for E2 (as can be observed when comparing the first and fourth rows in Fig.~\ref{fig:Cor_real_SC_sum}), however the final reconstruction is slightly worse (see last row in Fig.~\ref{fig:Cor_real_SC_sum});} 
{the same happens for increasing $C_{\rm ani}$ at fixed $S$, but the effect is much smaller, as indicated by the rightmost column of this figure; this is likely because our mock galaxy catalogues have a small fraction of satellites ($\sim11\%$), so that the FoG effect is not particularly important.}


We next move on to estimator E3. Fig.~\ref{fig:quadrupole_SC_sum} shows the quadrupole moments of reconstructed galaxy catalogues, $\xi_2\left[{\bf x}^{(k)}\right](s)$, for the same $(S,C_{\rm ani})$ parameter combinations as in Fig.~\ref{fig:Cor_real_SC_sum}. The two grey dashed lines, which are the same in all subpanels, are respectively the quadrupole moments measured from the final galaxy catalogues at $z=0.5$ in real (upper) and redshift (lower) space, and as expected the former is zero on all scales probed here ($r\gtrsim10\Mpch$) while the latter is negative as a result of the Kaiser effect.

There are a few features in Fig.~\ref{fig:quadrupole_SC_sum} which are noticeable. First of all, \revise{we find rapid and monotonic convergences, with the solutions generally never requiring more than just a few iterations for all the parameter combinations. The convergence becomes slower for small smoothing lengths ($S=5\Mpch$), but the differences are minor.} 
Secondly, unlike for E2, here the choice of $S$ can have a significantly greater impact on the converged result of $\xi_2\left[{\bf x}^{(K)}\right]$: in the better scenarios, such as $(S,C_{\rm ani})=(9\Mpch,1.0)$, we can observe that $\xi_2\left[{\bf x}^{(K)}\right](s)\simeq0$ for \revise{$s\gtrsim20\Mpch$}, while in the less good cases, such as $(S,C_{\rm ani})=(15\Mpch,2.0)$ this can only be achieved at $s\gtrsim50\Mpch$. Third, overall speaking, if the smoothing length $S$ is too large, there is insufficient correction to make $\xi_2\left[{\bf x}^{(K)}\right]$ go to zero on all but the largest scales ($s\gtrsim50\Mpch$), while if $S$ is too small, the correction seems to `overshoot' and make $\xi_2\left[{\bf x}^{(K)}\right]$ positive. This can be reasonably explained, given that over-smoothing (i.e., a too large $S$) would lead to ${\bf\Psi}^{(k)}_{\rm S}$ values which are appropriate only for large scales and therefore the resulting corrections to galaxy coordinates are not enough on small scales, while in the case of under-smoothing (a too small $S$) the resulting values of ${\bf\Psi}^{(k)}_{\rm S}$ can be strongly affected by structures on very small scales, causing `too much' correction. Finally, for a specific $S$, varying $C_{\rm ani}$ between $1.0$ and $2.0$ does not seem to have a significant impact on the converged result of E3 (after four iterations, see the right column of Fig.~\ref{fig:quadrupole_SC_sum}).

Figure \ref{fig:monopole_SC_sum} is similar to Fig.~\ref{fig:quadrupole_SC_sum}, but shows the impact of $(S,C_{\rm ani})$ on estimator E4, i.e., $\xi_0\left[{\bf x}^{(k)}\right](s)/\xi_{\rm gg}(r)$. The convergence properties are \revise{comparable to the case of E3, with convergence achieved after two to four} iterations in all cases, and the observation in the cases of E2 and E3 that $C_{\rm ani}$ has a negligible effect holds here as well. The result is again sensitive to $S$, with a value of $S$ that is too small producing insufficient correction to bring E4 to $1.0$ on all scales, while an $S$ value that is too large causes an incorrect shape of E4 as a function of $s$ by deviating it from a constant value in $s$. Overall, we find that \revise{$S=8$-$9\Mpch$} is capable of bringing E4 closest to $1.0$ on all scales $s\gtrsim20\Mpch$.

Very reassuringly, in general, for combinations $(S, C_{\rm ani})$ that bring $\xi_2\left[{\bf x}^{(k)}\right]$ closer to zero down to small $s$ values, the corresponding $\xi_2\left[{\bf x}^{(k)}\right]/\xi_{\rm gg}(r)$ curves are also close to $1.0$, which suggests that the reconstruction can get the two correct simultaneously (as it is expected to).

To summarise, we find that
\begin{enumerate}
    \item \revise{estimator E2 prefers a larger smoothing length $S$, and the value of $C_{\rm ani}$ is not as important;}
    \item compared with the E2 estimator, E3 and E4 are more sensitive to $S$, and disfavour either very large or very small values of $S$;
    \item the key objective of the reconstruction algorithm, namely to accurately remove the RSD effects (or equivalently to bring E3 to $0$ and E4 to $1$), \revise{can be achieved for $S\sim8$-$9\Mpch$ and $C_{\rm ani}=1.0$}.
\end{enumerate} 
These have motivated us to choose $9\Mpch$ as the optimal value for $S$ (for galaxy number density $n_{\rm g}=3.2\times10^{-4}\left[\Mpch\right]^{-3}$). As for $C_{\rm ani}$, given its weak impact on all estimators, we opt for the simple choice by setting its default value to $1.0$. \revise{This simplifies our pipeline as it now adopts isotropic smoothing, but note that for other $n_{\rm g}$ values this needs to be checked separately}.

The results for E1 are very similar to E2, and to avoid getting this paper too heavy on technical details, we refrain from showing them here.

\subsection{Galaxy bias parameter $b^{(k)}$}

\begin{figure*}
\centering
	\includegraphics[width=1\textwidth]{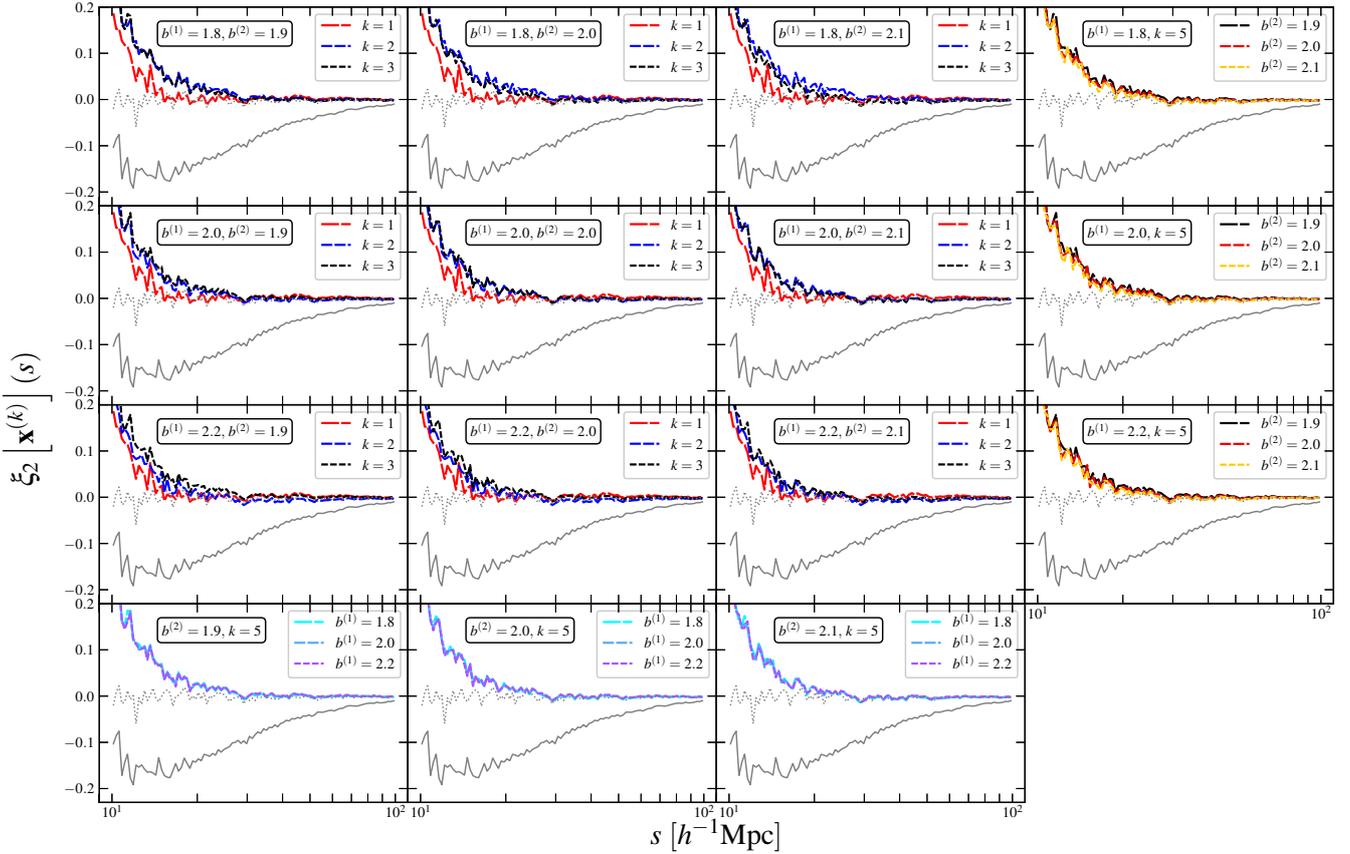}
	\vskip -.0cm
    \caption{{[{\it Colour Online}]} Estimator E2,  $\xi_2\left[{\bf x}^{(k)}\right](s)$ for the set of updated galaxy coordinates $\bf x^{\it k}$ after $\it k$ th iterations for different combinations of galaxy bias parameter $b^{(k)}$. The bias, $b^{(k)}$, is the one used in the $k+1$ th iteration of the reconstruction method. The values of the $b^{(1)}$ and $b^{(2)}$ bias parameters are shown in the label of each plot, while $b^{(0)}=2.2$ is the same for all panels and thus is not shown.
    The first three rows contains the tests that use a fixed $b^{(1)}=1.8, 2.0$ and $2.2$ while each column gives the results for a fixed $b^{(2)}= 1.9, 2.0$ and $2.1$ respectively. So the upper left corner contains $3 \times 3$ subplots and each of the subplot represents a unique combination of $b^{(1)}$ and $b^{(2)}$, showing the variation of E2 as the iteration number increases. {The sparseness of the dahses lines increases with $k$, $b^{(1)}$ or $b^{(2)}$ respectively in the three different regions (see the legends).}
    The {grey dotted and solid} lines have the same meaning as in Fig.~\ref{fig:quadrupole_SC_sum}. Each of the subpanels on the rightmost side show how varying $b^{(2)}$ for a fixed $b^{(1)}$ affects the reconstruction outcome, while  each of the subpanels at the bottom illustrate the effect of varying $b^{(1)}$ for fixed $b^{(2)}$ values.
    }
    \label{fig:quadrupole_BABBBC_sum}
\end{figure*}

\begin{figure*}
\centering
	\includegraphics[width=1\textwidth]{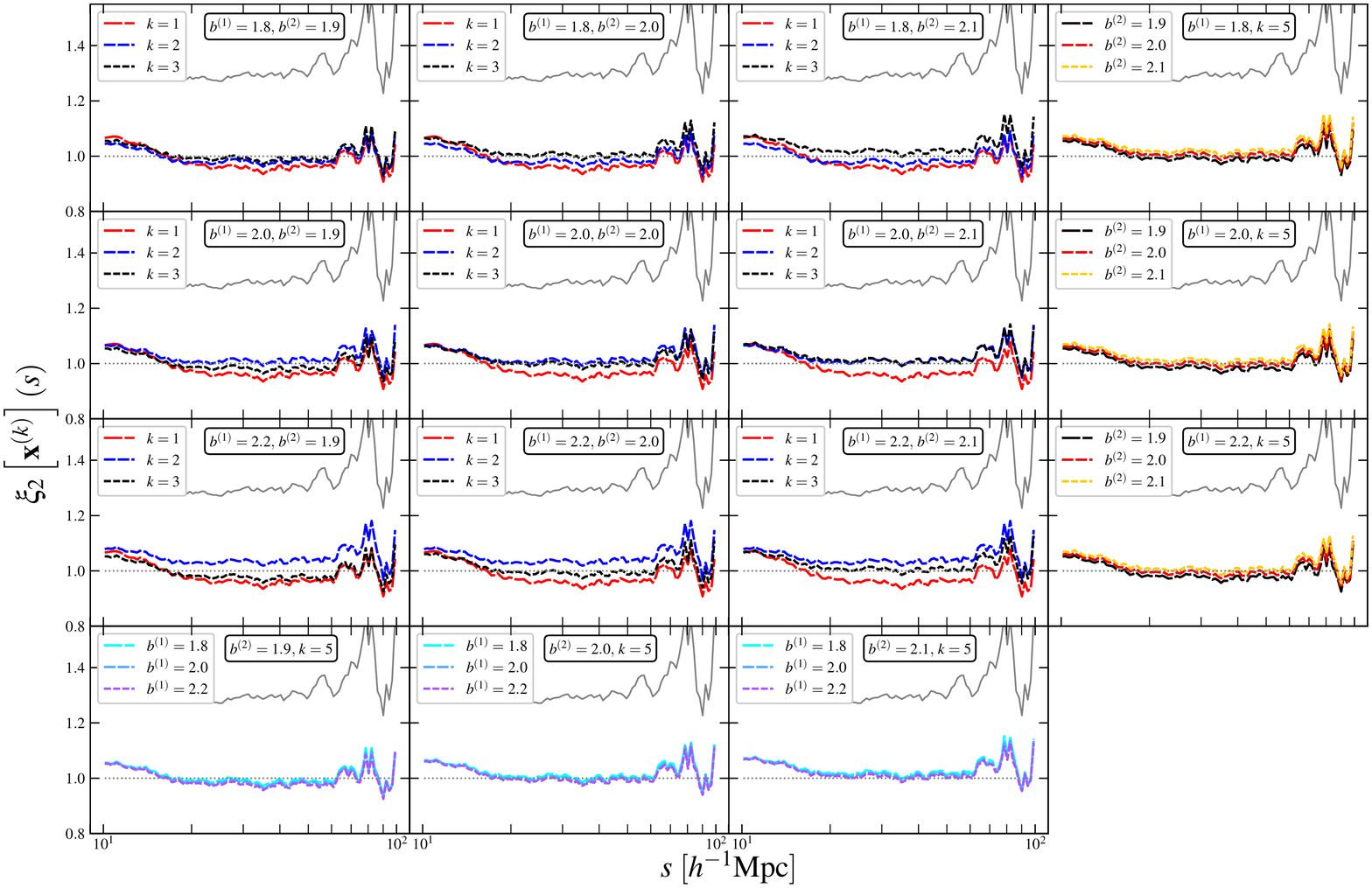}
	\vskip -.1cm
    \caption{{[{\it Colour Online}]} The same as Fig.~\ref{fig:quadrupole_BABBBC_sum}, but for estimator E4, $\xi_0\left[{\bf x}^{(k)}\right](s)/\xi_{\rm gg}(r)$. The straight {dotted} and the wiggly {solid} grey lines are the same in all subpanels: the former is the constant $1.0$ to guide the eyes, while the latter is the ratio between the multipole moments measured from the redshift- and real-space galaxy catalogues at $z=0.5$. 
    }
    \label{fig:monopole_BABBBC_sum}
\end{figure*}

Next, we explore the impact on reconstruction of the linear galaxy bias parameter, $b^{(k)}$. As mentioned above, this parameter is used to convert a nonlinear galaxy density field to a nonlinear matter density field, since it is the latter that enters the reconstruction equation \citep{Shi:2017gqs,Birkin:2018nag}. In the $\Lambda$CDM scenario, linear bias is time dependent but scale independent on large and linear scales. Therefore, given that we work at a fixed redshift, $z=0.5$, we simply take $b^{(k)}$ as a constant number.

While the linear galaxy bias is a physical parameter, we do not necessarily know its value accurately as this depends on the galaxy population. This is especially true in observations, where we do not even have precise knowledge of the cosmological parameters. As a result, by trying different values of $b^{(k)}$, we can test whether the exact value adopted is important -- if yes, then the reconstruction can be used to determine this value; if not, then not precisely knowing its value would not impact the reconstruction outcome strongly. 

We do, on the other hand, allow the bias $b^{(k)}$ to vary between the different iterations in the reconstruction process -- and there is a reason for this. Usually, when speaking about galaxy bias, one refers to the bias in real space, $\delta^r_{\rm g}\equiv b\delta_{\rm m}$, where $\delta^r_{\rm g}$ and $\delta_{\rm m}$ are the density contrasts of galaxies and matter in real space, respectively. However, at a given iteration of reconstruction, especially when $k=0$, what we have are not exactly the galaxy coordinates in real space but some approximations (for $k>0$), or their coordinates in redshift space (for $k=0$). Therefore, to convert the galaxy density field $\delta^s_{\rm g}$ or $\delta^{(k)}_{\rm g}$ to the real-space matter density field, an additional bias is needed and this additional bias depends on how much deviation $\delta^s_{\rm g}$ or $\delta_{\rm g}^{(k)}$ has with respect to the real-space galaxy density field, $\delta^r_{\rm g}$. One can argue that $b^{(0)}$ should be the largest because the additional bias will apply to $\delta^{s}_{\rm g}$ which differs most from $\delta^r_{\rm g}$, while for $k>0$ the additional bias correction required should decrease as $\delta^{(k)}_{\rm g}$ gets closer to $\delta^r_{\rm g}$\footnote{This additional bias is also one of the reasons why the tests do not use the galaxy bias value directly measured from simulations, $b_{\rm sim}$. However, just for completeness, we report the simulation result here -- $b_{\rm sim}=1.956$, which is obtained as the ratio of the galaxy auto correlation function $\xi_{\rm gg}(r)$ and galaxy-matter cross correlation function $\xi_{\rm gm}(r)$ at $r\gtrsim5\Mpch$, both measured from the simulation using the {\sc cute} code \citep{Alonso2012}.}.

For this reason, in the tests here we sample a $2\times3\times3$ grid of the parameter space, with $b^{(0)}\in\left\{2.3, 2.0\right\}$, 
$b^{(1)}\in\left\{1.8, 2.0, \revise{2.2}\right\}$ and $b^{(2)}\in\left\{1.9, 2.0, 2.1\right\}$. For further iterations ($k\geq3$), we simply fix $b^{(k)}=2.0$ because, as we shall see shortly, while differences can be spotted between $b$ being $1.8, 2.0$ and \revise{$2.2$}, it is very mild between $1.9, 2.0$ and $2.1$. The other reconstruction and physical parameters are fixed to $S=9.0\Mpch$, $C_{\rm ani}=1.0$ and $f=0.735$ for the $b^{(k)}$ tests presented in this subsection. 

The test results for the E3 estimator (the quadrupole moment) are presented in Fig.~\ref{fig:quadrupole_BABBBC_sum}, which demonstrate how marginal the differences are between the different choices of $b^{(k)}$. For all curves in this figure we have fixed $b^{(0)}=2.3$ because we have checked that the results for $b^{(0)}=2.0$ are almost identical. In the block of $3\times3$ panels at the top left corner, each row has a fixed $b^{(1)}$ and each column has a fixed $b^{(2)}$; the legend for each curve not only shows the corresponding values of $b^{(k)}$ but also indicates the current iteration number $k$: for example, `$b^{(2)}=1.9$' in the top left panel means that this is the result {\it after} iteration $k=3$, with $b^{(0)}=2.3$, $b^{(1)}=1.8$ and $b^{(2)}=1.9$\footnote{Note that $b^{(k)}$ is applied to the galaxy density field $\delta^{(k)}_{g, \rm s} $ {\it in} the $(k+1)$th iteration, while its effect can be seen in $\xi_{0,2}\left[{\bf x}^{(k+1)}\right]$, i.e., {\it after} the $(k+1)$th iteration, only. Therefore, the curve labelled as `$b^{(2)}$' is actually the result after three iteration. The convention of using $k$ is indicated in Fig.~\ref{fig:flowchart}. }, and so on. In all cases, we find that without iterations ($k=0$) the estimator E3 of the \revise{unreconstructed} galaxy catalogue is visibly nonzero at $s\lesssim60\Mpch$, while it rapidly converged to $0$ at $s\gtrsim15\Mpch$ after one or two iterations. The precise values of $b^{(1)}$, in the range of $[1.8,\revise{2.2}]$, and $b^{(2)}$, in the range of $[1.9,2.1]$, have little impact on the final converged results. 

Fig.~\ref{fig:monopole_BABBBC_sum} has the same layout as Fig.~\ref{fig:quadrupole_BABBBC_sum}, but shows the results for estimator E4, or $\xi_0\left[{\bf x}^{(k)}\right](s)/\xi_{\rm gg}(r)$. This plot again indicates that the exact choices of $b^{(k)}$ have a relatively small effect, with a larger $b^{(1)}$ tending to `undo' the improvement by iterative reconstruction (cf.~blue curves in the second and third rows), while further iterations tending to restore that improvement (black curves in the same panels). \revise{In all cases, having $b^{(k)}=2.0$ for $k>0$, makes E4 close to $1.0$, which is not surprising given that $2.0$ is close to the linear bias value measured from the simulations, $1.956$.} 

\begin{figure*}
\hspace*{-1.2cm}
	\includegraphics[width=\textwidth]{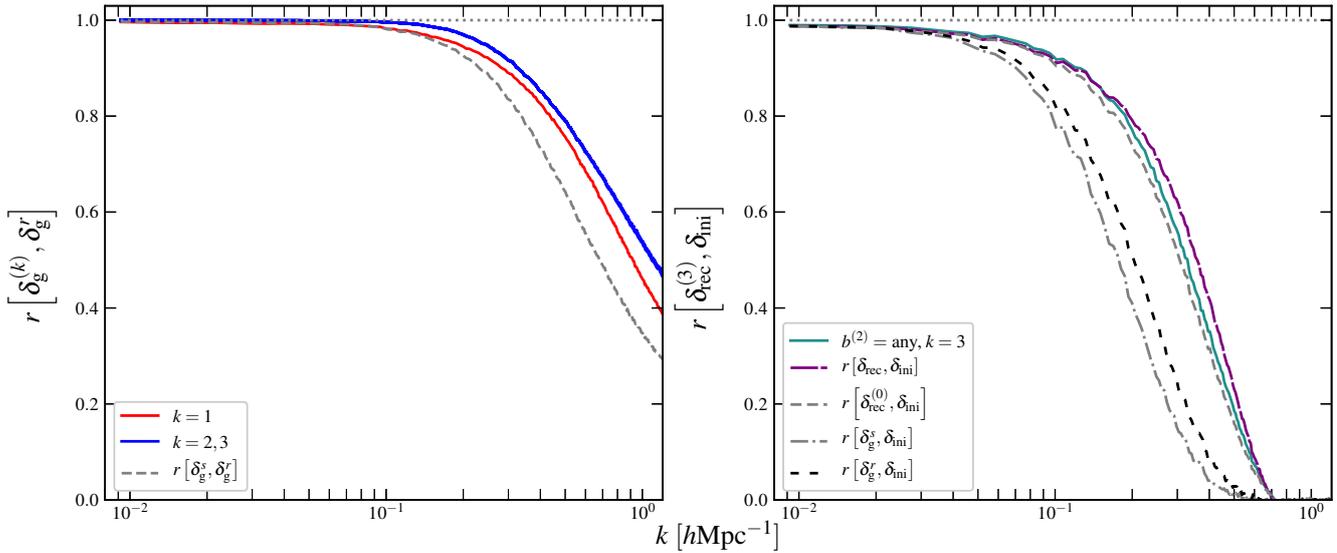}
 	\vskip -.2cm
    \caption{{[{\it Colour Online}]} {\it Left panel}: The estimator E2, $r\left[\delta^{(k)}_{\rm g},\delta_{\rm g}^r\right]$ after one (\revise{red} solid line {in the middle}), \revise{two and three (twelve blue solid lines, {which are indistinguishable and appear as the single thick solid line on the top})} iterations where the value of $b^{(k)}$ is allowed to vary across the different iterations; in all cases we have used $b^{(0)}=2.2$, $b^{(1)}\in\left\{1.8,2.0,2.2\right\}$, and $b^{(2)}\in\left\{1.9, 2.0, 2.1\right\}$. The blue line is actually 12 overlapping curves, of which 3 have $k=2$ and $b^{(1)}=1.8,2.0 \ \rm{and} \ 2.2$, and a further 9 lines which have $k=3$ and correspond to all possible combinations of $b^{(1)}$ and $b^{(2)}$. {\it Right panel}: The same as the left panel, but for estimator E1, $r\left[\delta^{(k)}_{\rm rec},\delta_{\rm ini}\right]$. Only the results after three iterations are shown (nine green solid lines, {which are indistinguishable and appear as the single solid line on the right}). The {black sparsely dashed (second from the left) and grey dot-dashed (leftmost)} lines are the correlation coefficients of the initial matter density field $\delta_{\rm ini}$ with the real- and redshift-space galaxy density fields respectively; the {purple long dashed (rightmost) and grey short dashed (left to the solid line)} lines are the correlation coefficients of $\delta_{\rm ini}$ with the reconstructed matter density field from the real- and redshift-space galaxy density field (no iteration in the latter case) respectively.}
        \label{fig:Cor_real_B}
\end{figure*}

In the left panel of Fig.~\ref{fig:Cor_real_B}, we present the $b^{(k)}$ test results for estimator E2. The grey dashed curve at the very bottom is $r\left[\delta^{s}_{\rm g},\delta^{r}_{\rm g}\right]$ or equivalently $r\left[\delta^{(0)}_{\rm g},\delta^{r}_{\rm g}\right]$, and the black solid curve immediately above that is $r\left[\delta^{(1)}_{\rm g},\delta^{r}_{\rm g}\right]$ with $b^{(0)}=\revise{2.2}$, which indicates that the first iteration substantially improves the \revise{correlation} between the reconstructed and the real-space galaxy fields. 
Finally, \revise{the top blue curve actually represents a bunch of 12 lines which are so close to each other that they are indistinguishable by eye: these include three lines after the second iteration with $b^{(1)}=1.8,2.0,2.2$, and 9 lines after one further iteration, with $b^{(2)}=1.9, 2.0, 2.1$. Together these show that after two iterations the results have converged well.} Again, for this estimator we find a weak dependence of the converged result on the values of $b^{(k)}$. 

The right panel of Fig.~\ref{fig:Cor_real_B} presents the results for E1, in which \revise{the black and the grey} solid curves are respectively $r\left[\delta^{r}_{\rm g},\delta_{\rm ini}\right]$ and $r\left[\delta^{s}_{\rm g},\delta_{\rm ini}\right]$ -- the cross correlations between the initial matter density field $\delta_{\rm ini}$ and the (nonlinear) galaxy density fields from the real- and redshfit-space galaxy catalogues (with no iterations in both cases). The \revise{purple and grey} dashed lines denote, respectively, $r\left[\delta_{\rm rec},\delta_{\rm ini}\right]$ and $r\left[\delta^{(0)}_{\rm rec},\delta_{\rm ini}\right]$ -- the cross correlations between the initial matter density field and the reconstructed matter density fields from the real- and redshfit-space galaxy catalogues (again with no iteration in the latter case). In between the two dashed lines are a bunch of \revise{$12$} green solid lines -- indistinguishable by eye -- which show the reconstruction results after three iterations for different combinations of $b^{(0,1,2)}$. The iterative RSD reconstruction improves the reconstruction of the initial density field on all scales, while there is still some residual RSD effect that it fails to remove.

\begin{figure*}
\centering
 	\includegraphics[width=0.98\textwidth]{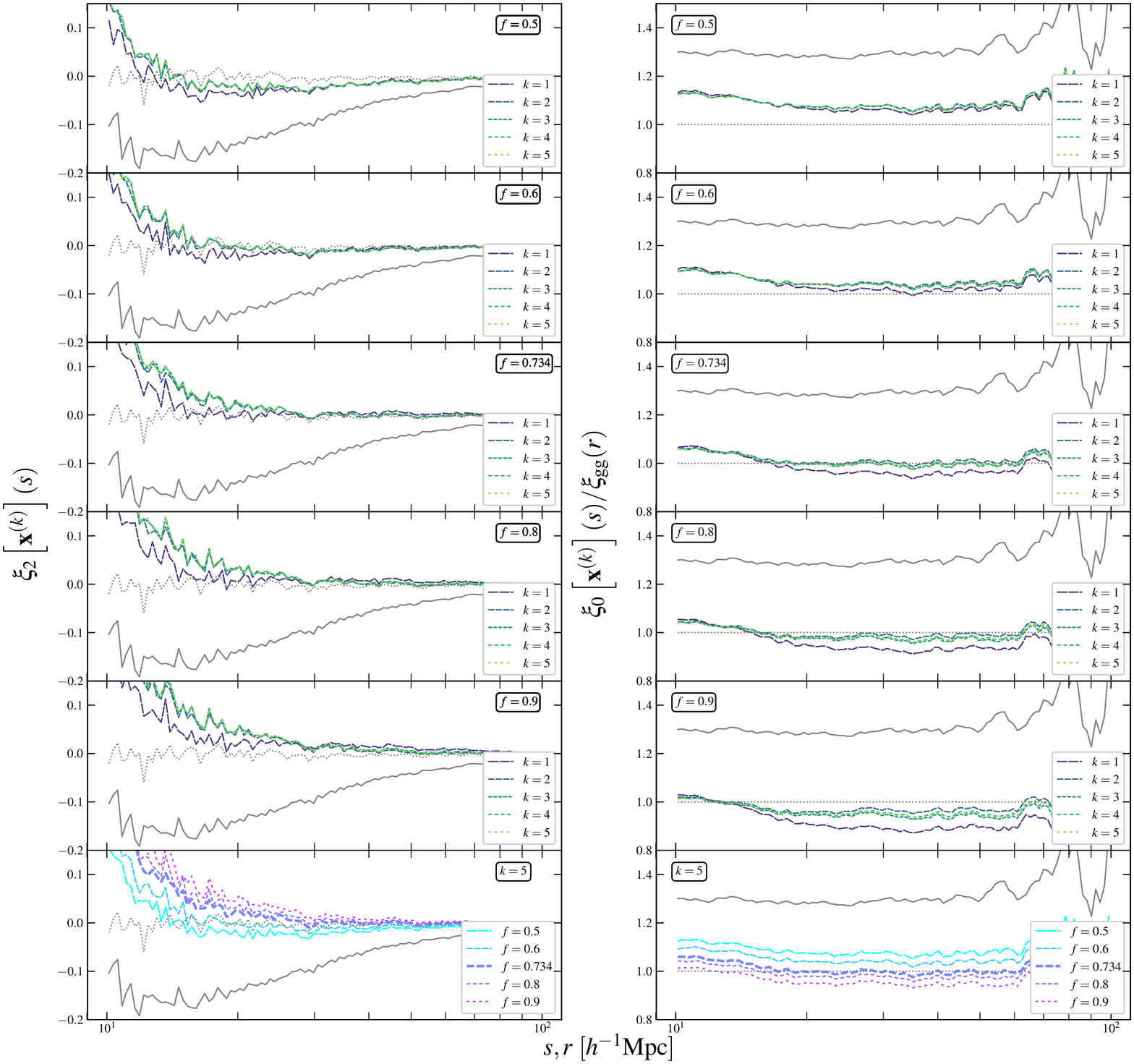}
     \caption{{[{\it Colour Online}]} {\it Left panels}: The estimator E3, $\xi_2\left[{\bf x}^{(k)}\right](s)$ for $f=0.5$, $0.6$, $0.735$ (theoretical value), $0.8$ and $0.9$ in the first five rows after $k$th iteration, respectively plotted in {dashed lines with thier sparsenesses increasing with $k$ (see the legends)}. The grey {dotted} curve (close to constant zero) is the quadruopole measured from the real-space galaxy catalogue, while the grey {solid} curve is that measured from the redshift-space catalogue before reconstruction. The last row compares the results from using the different $f$ values after the fifth iteration, {with the sparsenesses of the curves increasing with $f$ (see the legends)}. {\it Right panels}: the same as the left, but for estimator E4, $\xi_0\left[{\bf x}^{(k)}\right](s)/\xi_{\rm gg}(r)$. The grey {solid} curve is $\xi_0(s)/\xi_{\rm gg(r)}$, i.e., result before reconstruction, and the horizontal grey {dotted} line is $1$.
     }
     \label{fig:quadrupole_monopole_f_sum}
 \end{figure*}

\subsection{Linear growth rate $f$}
\label{sec:f}

Finally, we have tested the effect of the linear growth rate, $f$, in the reconstruction result, using a range of values between $0.5$ and $0.9$. In our reconstruction algorithm, the size of $f$ determines how much correction is applied to the redshift-space coordinates of galaxies -- a $f$ value that is too large will make the coordinates over-corrected and vice versa. Therefore, we expect that there is a limited range of $f$ which would lead to sensible reconstruction result.

We have adopted the following values of the other parameters -- $S=9\Mpch$, $C_{\rm ani}=1$, $b^{(0)}=2.3$ and \revise{$b^{(k>0)}=1.9$} -- in all the tests mentioned in this subsection. The left panels of Fig.~\ref{fig:quadrupole_monopole_f_sum} show the estimator E3, $\xi_2\left[{\bf x}^{(k)}\right](s)$, respectively for $f$ from $0.5$ to $0.9$ (the first five rows); the last row compares the results from using the different $f$ values after the fifth iteration. As anticipated above, we confirm that using $f$ values which are too small ($f=0.5,0.6$) leads to incomplete elimination of the quadrupole at \revise{$s\gtrsim30\Mpch$}. Likewise, when the adopted value of $f$ (e.g., $f=0.8,0.9$) is larger than the correct one, $f=0.735$, the quadrupole is over-corrected and becomes slightly positive between \revise{$30$} and $40\Mpch$, \revise{though both effects are weak}. 

The right panels of Fig.~\ref{fig:quadrupole_monopole_f_sum} are the same as the left panels, but for the estimator E4, $\xi_0\left[{\bf x}^{(k)}\right](s)/\xi_{\rm gg}(r)$. The behaviour is broadly consistent with what we have found for E3: when $f$ is too small, the reconstruction, even after convergence, is unable to completely remove the RSD effect and bring E4 to unity, while using a value of $f$ that is too large over-corrects the monopole by making it smaller than the real-space galaxy correlation function. 

There findings seem to suggest that we can use the reconstruction algorithm to place a \revise{constraint} on $f$. However, recall that in linear RSD studies there is a degeneracy between $f$ and the linear galaxy bias $b$. Schematically, the velocity divergence $\theta$ and linear matter perturbation $\delta$ are related by $\theta=aHf\delta$ whereas the galaxy density field $\delta_{\rm g}=b\delta$, so that $\theta\propto(f/b)\delta_{\rm g} \equiv \beta\delta_{\rm g}$. Although $f$ and $b$ enter the reconstruction pipeline at different places, this degeneracy will persist in the following way: the galaxy density contrast $\delta_{\rm g}$ is first divided by $b$ to get the matter density field; the latter is used to find the displacement field ${\bf\Psi}$; while the reconstruction solves a nonlinear equation to calculate ${\bf\Psi}$ from $\delta$, for large scales the two quantities satisfy a linear relation to a good approximation, so that ${\bf\Psi}$ is `proportional' to $1/b$; then as ${\bf\Psi}$ is multiplied by $f$ according to Eq.~\eqref{eq:iterative_method} we get the $f/b$ dependence. As in the tests of this section we have fixed $b$, we expect the situation will get more complicated when we allow both $b$ and $f$ to be chosen without prior knowledge. Indeed, as we will see in the next section, using estimator E3 alone only gives a constraint on $\beta$.

\begin{figure*}
\hspace*{-1.2cm}
	\includegraphics[width=2\columnwidth]{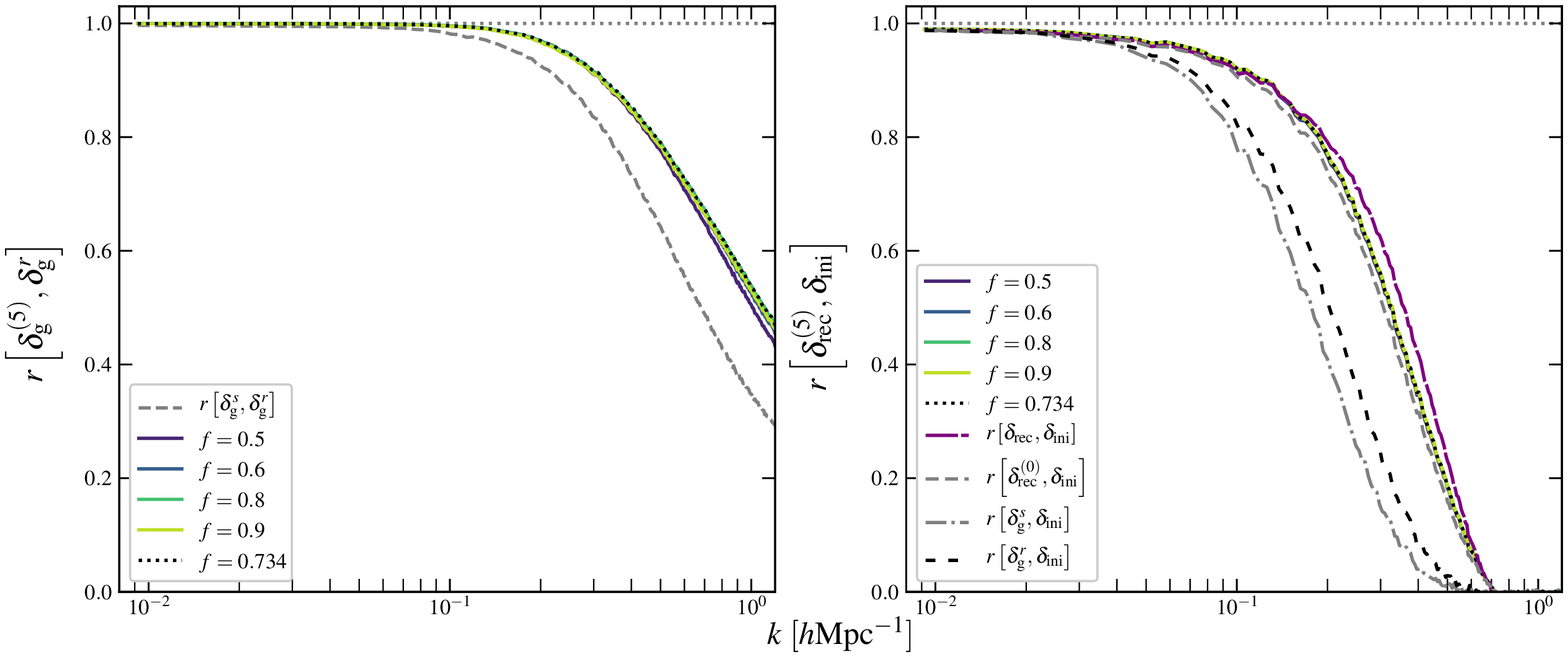}
	\vskip -.3cm
    \caption{{[{\it Colour Online}]} {\it Left panel}: estimator E2, $r\left[\delta^{(k)}_{\rm g},\delta^r_{\rm g}\right]$ (the correlation between the reconstructed and real-space galaxy density fields), for different values of $f$ after five iterations of the reconstruction procedure. The black dotted line represents $f=0.734$ and other $f$ values are shown by coloured solid lines as indicated in the legend {(note that all the coloured solid lines but the case for $f=0.5$ -- which is the lowest of them -- are indistinguishable)}. The grey dashed curve is the correlation between the redshift-space galaxy density field before reconstruction, $\delta^{(0)}_{\rm g}$, and the real-space galaxy density field $\delta^r_{\rm g}$. {\it Right panel}: the same as the left, but for estimator E1, $r\left[\delta^{(k)}_{\rm rec},\delta_{\rm ini}\right]$ (the correlation between the reconstructed and initial matter density fields). The {black sparsely dashed (second from the left)} and grey {dot-dashed (leftmost)} lines are the correlation coefficients of the initial matter density field $\delta_{\rm ini}$ with the real- and redshift-space galaxy density fields respectively; the {purple long dashed (rightmost)} and grey {dashed (third from the left)} lines are the correlation coefficients of $\delta_{\rm ini}$ with the reconstructed matter density field from the real- and redshift-space galaxy density field (no iteration in the latter case) respectively. {Note that all the coloured solid lines (for different $f$ values) are indistinguishable from each other, and the black dotted line, which is for $f=0.734$, is on top of them (they appear as a single line (secon from the right)).}} 
        \label{fig:Cor_real_F}
\end{figure*}

The results for estimators E2 (correlation between the reconstructed and real galaxy density fields) and E1 (correlation between the reconstructed and initial matter density fields) are presented respectively in the left and right panels of Fig.~\ref{fig:Cor_real_F}. Only results after five iterations are shown. The impact of using different values of $f$ on these estimators is again mild. From the left panel we can see that for all $f$ values the correlation is substantially improved compared with the case of no reconstruction (dashed line), and whereas using small values of $f$ can slightly degrade the performance, using $f=0.9$ gives practically indistinguishable result from the case of $f=0.735$. The same happens to estimator E1 which is shown in the right panel, for which the difference between different $f$ values is even smaller. Overall, we find that, for our chosen galaxy number density, $n_{\rm g}=3.2\times10^{-4}\left[\Mpch\right]^{-3}$, the RSD effect on reconstruction is small, e.g., by comparing $r\left[\delta_{\rm rec},\delta_{\rm ini}\right]$ with $r\left[\delta^{(0)}_{\rm rec},\delta_{\rm ini}\right]$, and the improvement of the iterations is even smaller, e.g., by comparing the black dotted line with $r\left[\delta^{(0)}_{\rm rec},\delta_{\rm ini}\right]$ \revise{(the grey dashed line)}. 

\section{An application of the method}
\label{sect:application}

\begin{figure*}
	\includegraphics[width=\textwidth]{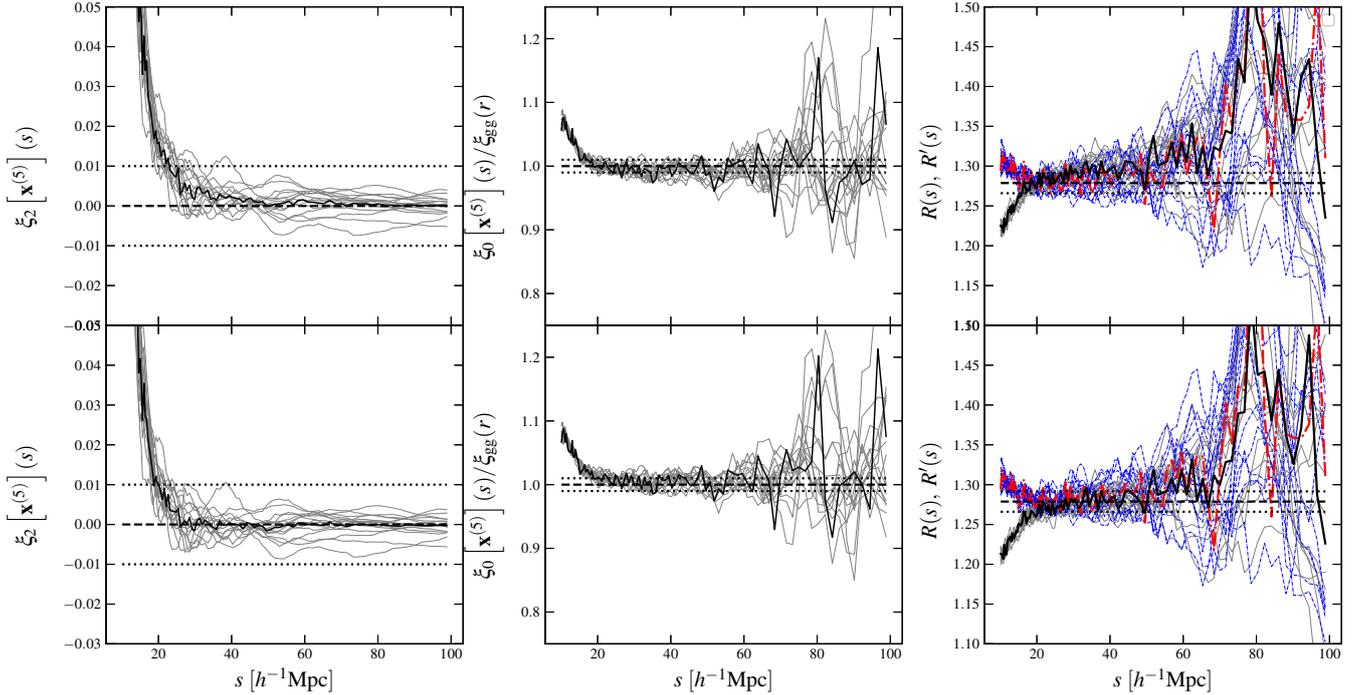}
    \caption{{[{\it Colour Online}]} The quadrupole ($\xi_2\left[{\bf x}^{(5)}\right](s)$; left column), monopole ($\xi_0\left[{\bf x}^{(5)}\right](s)/\xi_{\rm gg}(r)$; central column) and $R(s)\equiv\xi_0(s)/\xi_0\left[{\bf x}^{(5)}\right](s)$ (right column) of the reconstructed galaxy catalogues from the 15 realisations of mock redshift-space galaxy catalogues. The results for individual realisations are plotted as thin grey {solid} curves with their mean as thick black {solid} curves. In the right column, we also show the results of $R'(s)\equiv\xi_0(s)/\xi_{\rm gg}(r)$, the ratio between the redshift-space monopole and the real-space galaxy correlation function for comparison, with the thin blue {dash-dotted} lines showing the individual realisations and the thick red {dash-dotted} lines their mean. The top row shows the reconstruction result using theoretical values $f=0.734$ and $b=1.95$, while the bottom row shows the result using the best-fit value of $\beta=f/b$ obtained by minimising the derivation of $\xi_2\left[{\bf x}^{(5)}\right](s)$ from $0$ (see Section \ref{sect:application} for details), which corresponds to \revise{$\beta=0.356$}. The dotted horizontal lines in the left and central columns mark the $\pm0.01$ deviation from and $0$ and $1$ {, and in the right column they mark the $\pm\%1$ deviation from $R(s)$ calculated according to Eq.~(\ref{eq:monopole_ratio_practical}) with the theoretical value of $\beta$}.}
    \label{fig:15_combines}
\end{figure*}

\begin{figure*}
	\includegraphics[width=\textwidth]{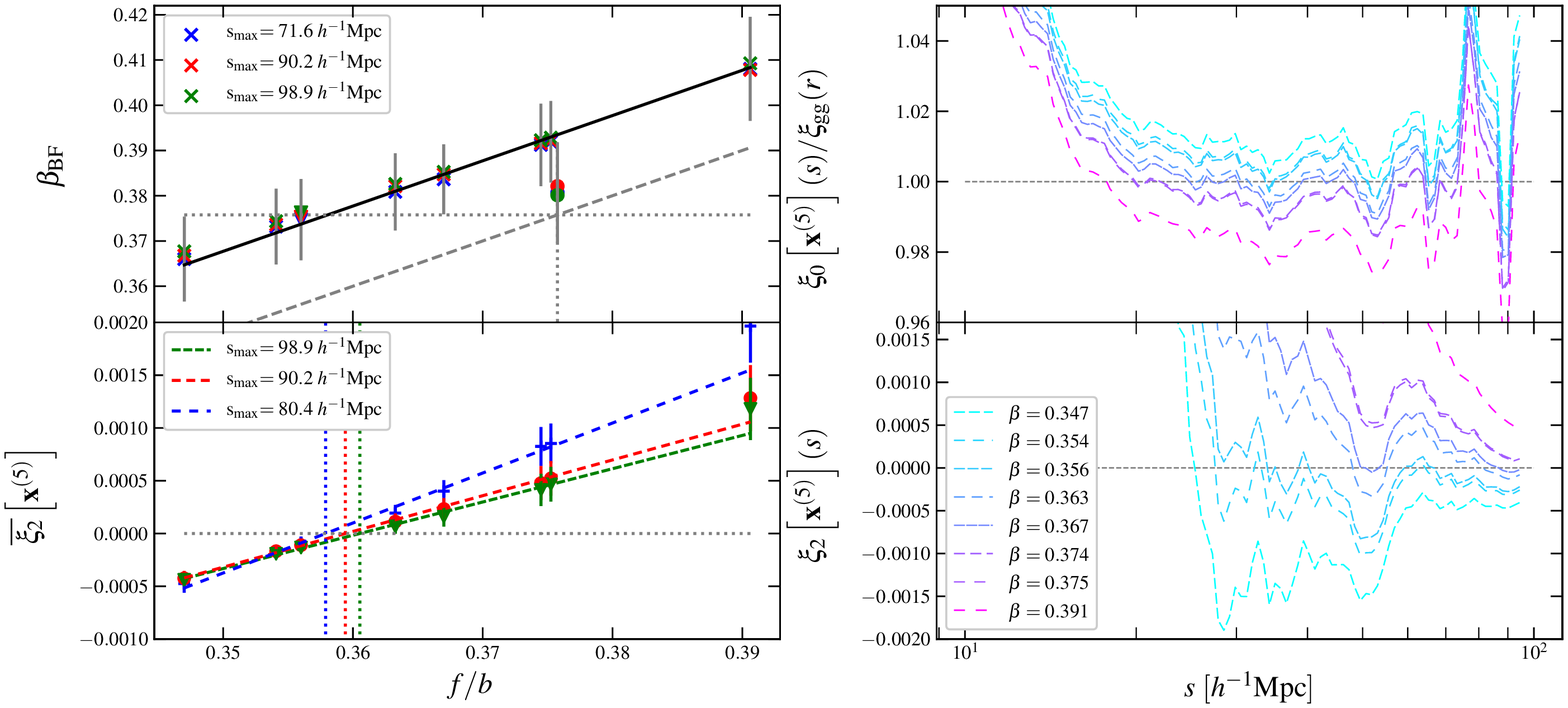}
    \caption{{[{\it Colour Online}]} {\it Upper left}: The best-fit $\beta_{\rm BF}$ values obtained using the estimator $R(s)$ and the reconstructed galaxy catalogues, for seven choices of input parameters $f$ and $b$ supplied to the reconstruction pipeline (crosses). $R(s)$ is measured as the weighted average between $s_{\rm min}=22.1\Mpch$ and $s_{\rm max}=98.9$ (green), $90.2$ (red) and $71.6\Mpch$ (blue) respectively, and the error bars are the standard deviations obtained from 15 realisations for the case of $s_{\rm max}=71.6\Mpch$ as the error bars for remaining $s_{\rm max}$ values are similar. The filled circles are the same, but obtained from the estimator $R'(s)$: since there is no reconstruction in this case, these data are plotted against the theoretical value $\beta=f/b=0.376$. \revise{The dashed line represents $\beta=f/b$, while the solid line is a line of slope $1.0$ to guide the eyes.} {\it Lower left}: $\bar{\xi}_2\left[{\bf x}^{(k)}\right]$, the constant value fitting $\xi_2\left[{\bf x}^{(k)}\right](s)$ between $s'_{\rm min}=39.4\Mpch$ and the three $s_{\rm max}$ values $s_{\rm max}=98.9$ (green, {densely dashed}), $90.2$ (red, {dashed}) and $80.4\Mpch$ (blue, {sparesly dashed}) respectively, for the same seven choices of input $(f,b)$. The vertical dotted lines mark the best-fit $\beta_{\rm BF}$ at which $\bar{\xi}_2\left[{\bf x}^{(k)}\right]$ is closest to zero, and the error bars are again the standard deviations of the 15 realisations. {\it Right panels}: the estimator E4 (upper) and E3 (lower) for the seven choices of reconstruction parameters $(f,b)$, with their corresponding values of $\beta$ indicated in the legend {and increase with the sparsenesses of the curves ($\beta$ increases from to the top (bottom) to the bottom (top) line in the upper (lower) right panels)}. Note that according to the lower left panel, the best value of $\beta$, amongst the seven choices, is $\beta=0.356$, and results for this choice are shown as triangles in the upper left panel.}
    \label{fig:Beta_result}
\end{figure*}

Having now acquired a more complete picture about the behaviour of the reconstruction algorithm and its dependence on the physical and technical parameters, we naturally would like to understand the potential of applying it to real galaxy data to extract cosmological information. Due to the limited scope of this paper, in this section we only attempt to explore this issue for an idealised setup -- reconstruction from mock galaxy catalogues in cubic simulation boxes -- and we shall comment on some of the complications when working with real data and leave detailed studies to future works.

As mentioned in the introduction, we wish the reconstruction algorithm to work in a ``self-calibration'' sense, such that the physical parameters, such as $f$ and $b$, can be determined during the reconstruction process itself. In the previous section we have seen that it is possible to `optimise' the method using estimator E3, $\xi_2\left[{\bf x}^{(K)}\right](s)$, and also that there is a degeneracy between $f$ and $b$ so that it is their combination $\beta\equiv f/b$ that matters. Here, we investigate whether the value $\beta$ can be accurately fixed by demanding that $\xi_2\left[{\bf x}^{(K)}\right](s)$ vanishes for the range of scales in which the iterative reconstruction method works. 

In linear perturbation theory, the integration of Kaiser formula \citep{Kaiser1987} to relate the real- and redshift-space galaxy density contrasts,
\begin{equation}
\label{Kaier}
\delta^{s}_{\rm g}(s, \mu) = \left(1 + \beta\mu^2\right)\delta^{r}_{\rm g}(r), 
\end{equation}
where $\beta$ is the distortion parameter introduced above, 
and $\mu$ is the cosine of the angle between the galaxy pair separation vector $\bf s$ and the LOS direction, gives the monopole and quadrupole moments of the galaxy correlation function in redshift space  \citep{Hamilton1993}
\begin{align}
\label{eq:monopole_ratio_theoretical}\xi_0(s) &= \left(1 + \frac{2}{3}\beta + \frac{1}{5}\beta^2\right) \xi_{\rm gg}(r),\\ 
\xi_2(s) &= \left(\frac{4}{3}\beta + \frac{4}{7}\beta^2\right)\left[\xi_{\rm gg}(r) - \bar{\xi}_{\rm gg}(r)\right],
\end{align}
where $\bar{\xi}_{\rm gg}(r)$ is the average correlation function with $r$:
\begin{equation}
\bar{\xi}_{\rm gg}(r) \equiv 3r^{-3}\int_{0}^{r}\xi_{\rm gg}(r^{\prime})r^{\prime 2}dr^{\prime}.
\end{equation}

In practice, while $\xi_{0,2}$ can be directly measured from observations, these equations cannot be used to infer $\beta$ because the real-space correlation function $\xi_{\rm gg}(r)$ is not observable. 
However, it is possible to combine these equations to estimate $\beta$ from the ratio of redshift-space quadrupole and monopole moments:
\begin{equation}\label{eq:quadrupole_ratio_estimator}
    \frac{\xi_2(s)}{\xi_0(s)-\bar{\xi}_0(s)} = \frac{\frac{4}{3}\beta+\frac{4}{7}\beta^2}{1+\frac{2}{3}\beta+\frac{1}{5}\beta^2},
\end{equation}
with 
\begin{equation}
\bar{\xi}_{0}(s) \equiv 3s^{-3}\int_{0}^{s}\xi_{0}(s')s'^{2}ds'.
\end{equation}
The method has been applied to galaxy redshift surveys \citep{Peacock2001,Hawkins2003}, but studies using simulation data show that it is difficult to accurately recover the actual value of $\beta$ \citep[e.g.,][]{Cole1994,Hern2019}, while the estimator ${\xi_{0}(s)}/{\xi_{\rm gg}(r)}$ gives better measurement of $\beta$. One reason for this is that the estimator given in Eq.~(\ref{eq:quadrupole_ratio_estimator}) suffers from the often noisier measurement of $\xi_2(s)$. Furthermore, while Kaiser effect only takes account for the coherent linear motion of galaxies, the non-linearity induced by FoG effect could have a non-negligible impact on the quadrupole down to $s\sim50\Mpch$ \citep{Cole1994}, which invalidates the linear assumption in Eq.~(\ref{eq:quadrupole_ratio_estimator}). Thus, using only the monopole moment in extracting information according to linear theory could be desirable over the traditional method Eq.~(\ref{eq:quadrupole_ratio_estimator}): even though this requires knowledge of $\xi_{\rm gg}(r)$ which is unobservable, we have seen that reconstruction gives $\xi_0\left[{\bf x}^{(K)}\right](s)$ which can be used as an approximation to $\xi_{\rm gg}(r)$. 

In order to more quantitatively assess the accuracy of this approximation, we have measured $\xi_{2,0}\left[{\bf x}^{(K)}\right](s)$ from the 15 realisations of mock galaxy catalogues, and the results are shown, respectively, in the upper left and upper middle panels of Fig.~\ref{fig:15_combines}. There we have adopted the theoretical values of $f=0.735$ and $b=1.95$, so that $\beta=f/b=0.376$, for reconstructions in all 15 realisations; the thin lines are the results from the individual realisations while the thick lines are their means. We can see that using the `correct' values of $f,b$ gives $\xi_{2}\left[{\bf x}^{(K)}\right](s)=0$ and $\xi_{0}\left[{\bf x}^{(K)}\right](s)/\xi_{\rm gg}(r)=1$ at \revise{$s\gtrsim20$}, with an accuracy of $\sim0.01$ and $\sim1\%$ respectively. This confirms that we can actually replace $\xi_{\rm gg}$ in Eq.~(\ref{eq:monopole_ratio_theoretical}) with $\xi_0\left[{\bf x}^{(K)}\right]$ as a way to estimate $\beta$:
\begin{equation}\label{eq:monopole_ratio_practical}
R(s) \equiv \frac{\xi_0(s)}{\xi_0\left[{\bf x}^{(K)}\right](s)} =  1 + \frac{2}{3}\beta(s) + \frac{1}{5}\beta^2(s).
\end{equation}
Note that in linear theory $\beta$ is a constant, but here we have kept the $s$-dependence because the estimator itself can fluctuate around the constant value from one $s$ bin to another.

To use the estimator $R(s)$, we must first specify the range of $s$, $\left[s_{\rm min}, s_{\rm max}\right]$, within which $R(s)$ is evaluated. We choose $s_{\rm min}, s_{\rm max}$ to ensure that $R(s)$ is approximately constant in that range, which gives \revise{$s_{\rm min}\simeq22\Mpch$}; for $s_{\rm max}$, we have checked three different values, 
\revise{$71.6\Mpch$, $90.2\Mpch$ and $98.9\Mpch$}: in principle, we expect that $R(s)$ is closer to a constant at larger $s$, but in practice the estimator becomes quite noisy and the uncertainty becomes large in that regime due to cosmic variance, so that we expect including larger $s$ bins should have a relative small impact on the overall fitting result. We divide $\left[s_{\rm min}, s_{\rm max}\right]$ into $N=10$ bins equally spaced in logarithmic scale and use a least chi-square method to find the best fit by minimising $\chi^2$
\begin{equation}
    \chi^2 = \frac{1}{N-1}\sum_{i=1}^{N} {\frac{(R_i - R)^2}{\sigma_i^2}},
\end{equation}
where $R\equiv1+\frac{2}{3}\beta_{\rm BF}+\frac{1}{5}\beta_{\rm BF}^2$ with $\beta_{\rm BF}$ the best-fit value of $\beta$, $R_i$ is the average value of $R(s)$ in the $i$th bin and $\sigma_i$ is the corresponding standard deviation \revise{of the 15 realisations}. 

In the upper right panel of Fig.~\ref{fig:15_combines}, we have compared the estimator in Eq.~(\ref{eq:monopole_ratio_practical}) with the estimator $R'(s)\equiv\xi_0(s)/\xi_{\rm gg}(r)$ from Eq.~(\ref{eq:monopole_ratio_theoretical}), in which the thin blue (grey) lines denote the latter (former) for the 15 individual realisations, while the thick red (black) line denotes the mean of them. This plot shows that at $s\gtrsim20\Mpch$ $R(s)$ and $R'(s)$ agree with each other at percent level. We have also checked the inferred $\beta_{\rm BF}$ values, following the method described in the previous paragraph, from these two estimators respectively, and found a good agreement too. Note that a small error in the estimator $R(s)$ or $R'(s)$ can be translated to a larger error in $\beta$ because of the $1$ in $1+\frac{2}{3}\beta+\frac{1}{5}\beta^2$. 

It is also interesting to note in the upper right panel of Fig.~\ref{fig:15_combines} that the reconstruction result $R(s)$ (thick black line) is slightly less noisier than the theoretical result $R'(s)$ (thick red line). The noise in $R'(s)$ is generated because the random phases of the galaxy field are changed by the RSD mapping from real space ($\delta^r_g({\bf r})$) to redshift space ($\delta^s_g({\bf s})$) -- i.e., $\delta^s_g({\bf s})$ is not simply a constant amplitude enhancement of $\delta^r_g({\bf r})$ -- so that when taking the ratio $\xi_0(s)/\xi_{\rm gg}(r)$ there is no perfect cancellation. The reconstruction works to revert this change of phases, but in practice this cannot be done perfectly, which means that, compared with the real-space galaxy field, the phases of the reconstructed galaxy field are `more similar' to those of the redshift-space galaxy field. 
Thus, by dividing the two function as in Eq.~(\ref{eq:monopole_ratio_practical}), there is a better cancellation of the phase effect. 

In practice, we do not know the theoretical values of $f$ and $b$ {\it a priori}, and it is interesting to know what happens to the inferred $\beta$ value if incorrect values of $f,b$ are used in the reconstruction. In \citet{Birkin:2018nag}, for example, it was found that the reconstruction result of the initial condition phases is not particularly sensitive to the value of the linear bias $b$ used. Here, we have checked this by running the reconstruction pipeline on the 15 realisations of mock galaxy catalogues, using 7 different $(f,b)$ combinations, including two which give very similar values of $\beta=f/b$. \revise{More explicitly, we have adopted a $3\times2$ grid with $f\in\left\{0.694, 0.712, 0.734\right\},  b\in\left\{1.96, 2.0\right\}$, and two points with $(f,b)=(0.75, 1.92)$ and $(0.734, 1.956)$ to enrich the result. The corresponding $\beta$ ranges from $0.347$ to $0.391$ passing through $0.354$, $0.356$, $0.363$, $0.367$, $0.374$ (theoretical value) and $0.375$.} 
In these tests, the other reconstruction parameters are fixed to $S=9\Mpch$ and $C_{\rm ani}=1.0$. 


From the reconstructed catalogues we then measure $R(s)$ and use them to find the best-fit $\beta_{\rm BF}$. In the upper left panel of Fig.~\ref{fig:Beta_result} we plot $\beta_{\rm BF}$ against $f/b$ for these points as crosses; for each $f/b$ value, there are three $\beta_{\rm BF}$ values with different colours, respectively obtained by using $s_{\rm max}=98.9\Mpch$ (green), \revise{$90.2\Mpch$} (red) and \revise{$71.6\Mpch$} (blue). We make the following observations from this plot:
\begin{enumerate}
    \item $\beta_{\rm BF}$ versus $f/b$ falls nicely on a straight line with slope $1$ (the \revise{solid} line). \revise{However, note that the relation $\beta_{\rm BF}=f/b$ (the dashed line) does not exactly hold: we have checked that the actual value of $\beta_{\rm BF}$ depends on the $s_{\rm min}$ used, decreasing as smaller $s_{\rm min}$ is used; the results here are for $s_{\rm min}=22.1\Mpch$.} The trend suggests that, roughly speaking, the reconstruction simply spits out the "input" $\beta$ value, and therefore has no predictive power regarding $\beta$, unless we look beyond $R(s)$;
    
    \item the two $(f,b)$ pairs which give similar $\beta=f/b\approx0.375$ produce very similar $\beta_{\rm BF}$: this is a consequence of the $f$-$b$ degeneracy mentioned above; 
    
    \item for comparison, we include the best-fit $\beta_{\rm BF}$ values obtained using $R'(s)$ as \revise{filled circles}, using the same colour scheme. Because there is no reconstruction and therefore no input $(f,b)$, we simply plot the points at a fixed horizontal coordinate $f/b=0.376$ (the theoretical value). We note that $\beta_{\rm BF}$ obtained in this way is indeed $\sim1\%$ larger than the theoretical value, \revise{but this is well within the $1\sigma$ uncertainty indicated by the grey error bar; at this stage we can not accurately assess the impact on $\beta_{\rm BF}$ due to the small sample size or the simulation resolution, which is beyond the scope of the present work}. As mentioned above, a small (percent-level) error in $R'(s)$ leads to a larger error in $\beta_{\rm BF}$.
\end{enumerate}

This has motivated us to consider an alternative approach to find the best-fit value of $\beta$ from the reconstruction process, by demanding that $\xi_2\left[{\bf x}^{(K)}\right](s)$ is closest to zero on large, linear, scales. In practice, we quantify this using the mean value of $\xi_2\left[{\bf x}^{(K)}\right](s)$ between $s'_{\rm min}\simeq40\Mpch$ and \revise{$s_{\rm max}=98.9, 90.2, 80.4\Mpch$}, $\bar{\xi}\left[{\bf x}^{(K)}\right]$. In the lower left panel of Fig.~\ref{fig:Beta_result} we present $\bar{\xi}\left[{\bf x}^{(K)}\right]$ for the seven $(f,b)$ pairs as used above; the data points with different colours indicate the three choices of $s_{\rm max}$ and the dashed lines are the best-fit straight lines through the points. We note that as $f/b$ increases $\bar{\xi}\left[{\bf x}^{(K)}\right]$ goes from negative to positive, crossing zero at \revise{$f/b\approx0.36$}. This is indeed \revise{$\sim4\%$ smaller} than the theoretical value $f/b=0.376$, but note that {the resulting $\beta_{\rm BF}$ corresponding to this $f/b$ value (inferred from the upper left panel of Fig.~\ref{fig:Beta_result})} is actually closer to the best-fit value \revise{$\beta_{\rm BF}\approx0.38$} from $R'(s)$. \revise{This result is not surprising given that, as we shall see shortly, when using input value $f/b\simeq0.36$ the E4 estimator is very close to $1.0$ and $R(s)$ agrees better with $R'(s)$ at $s\gtrsim20\Mpch$ than the case where $f/b$ takes the theoretical value $0.376$.} We suspect that this small discrepancy is again due to the limited sample size or simulation resolution, and a more detailed investigation into this issue will be conducted in a forthcoming work with larger datasets.

The results of $\xi_2\left[{\bf x}^{(K)}\right](s)$ and of $\xi_0\left[{\bf x}^{(K)}\right](s)/\xi_{\rm gg}(r)$ for the seven $(f,b)$ pairs are respectively displayed in the upper right and lower right panels of Fig.~\ref{fig:Beta_result}. The $(f,b)$ pair whose corresponding $\beta$ value is closest to \revise{$0.38$} is \revise{$f/b=0.356$ (the third point from the left in the upper/lower left panels of Fig.~\ref{fig:Beta_result})}, and we can see that for this value $\xi_2\left[{\bf x}^{(K)}\right](s)$ is actually closest to zero (as expected) and $\xi_0\left[{\bf x}^{(K)}\right](s)/\xi_{\rm gg}(r)$ closest to $1.0$. This latter observation in particular indicates that by demanding $\bar{\xi}\left[{\bf x}^{(K)}\right]$ to be closest to zero for the best-fit $\beta$, we automatically obtain the $\xi_0\left[{\bf x}^{(K)}\right](s)$ that is closest to $\xi_{\rm gg}(r)$. 

To compare the performance of this new method to determine $\beta_{\rm BF}$, in the lower panels of Fig.~\ref{fig:15_combines} we have shown the same curves as in the upper panels, but this time $\beta=f/b=\revise{0.356}$. A careful visual inspection shows that the results of $\xi_2\left[{\bf x}^{(K)}\right](s)$ (lower left panel) and of $\xi_0\left[{\bf x}^{(K)}\right](s)/\xi_{\rm gg}(r)$ (lower middle panel) are actually closer to $0.0$ and $1.0$ respectively, than the cases shown in the upper panels of Fig.~\ref{fig:15_combines}, where the theoretical value $f/b=0.376$ is used. In the lower right panel we again find that $R(s)$ (black thick line) with $f/b=\revise{0.356}$ is nearly identical to $R'(s)$ (red thick line, which is the same as in the upper right panel) for $s\gtrsim20\Mpch$. \revise{This is why in the above we have adopted $s_{\rm min}\simeq22\Mpch$ for constraining $\beta_{\rm BF}$ using $R(s)$ and $R'(s)$.}

As highlighted above, in this section we have proposed to obtain the best-fit value of $\beta$ using a `self-calibration' of the observed (or mock) galaxy field, namely tune $\beta$ to minimise  $\overline{\xi}_2\left[{\bf x}^{(K)}\right]$. From lower left and upper right panels of  Fig.~\ref{fig:Beta_result} we have seen that the quadrupole of the reconstructed galaxy field changes slowly with $f/b$. 
The reconstruction method is a backward modelling approach, where one starts with a late-time observed galaxy field to infer information about the early-time density field, the peculiar velocity field and the cosmological model ($\beta$). This is different from the standard forward modelling approach where one starts with a given cosmological model, makes prediction and compares it against observations \citep[see, e.g.,][]{Sanchez:2016sas}. It is possible to adopt a hybrid approach. For example, one can start with a specific model, e.g., $\Omega_m,\sigma_8$, and predict the linear growth rate $f$ and linear matter correlation function at a $z$. The latter, together with $\xi_0\left[{\bf x}^{(K)}\right](s)$ as an approximation to $\xi_{\rm gg}(r)$, can be used to determine the linear bias $b$ and break its degeneracy with $f$. The former can then be checked against the $f$ determined as the ratio between the best-fit $\beta_{\rm BF}$ and the $b$ obtained in this way, and against the velocity power spectrum from the reconstructed galaxy catalogue. A detailed investigation of this, however, is beyond the scope of this work.

\section{Discussion and conclusions}
\label{sect:conclusions}

We have proposed and tested a new iterative scheme to reconstruct, simultaneously, the real-space {\it galaxy coordinates} at late times and the {\it initial matter} density field, from a given late-time redshift-space galaxy catalogue. The method builds on the nonlinear reconstruction algorithm developed by \citet{Shi:2017gqs,Birkin:2018nag}, taking into account the (linear) galaxy bias $b$ and redshift space distortions. It employs a number of (semi)-free parameters related to the smoothing of the galaxy field, $b$ and linear growth rate $f$. Since this is a new method and code, we have performed various checks not only to assess its reliability in obtaining physical results but also to understand how the reconstruction outcomes react to the specific choices of technical parameters.

The iterative reconstruction consists of continuous updates of the trial real-space galaxy coordinates based on improved knowledge of the large-scale displacement field ${\bf\Psi}$, until the results, as quantified by estimators E1 - E4\footnote{As we discussed, in practice only E3, the quadrupole of the reconstructed galaxy catalogue, can be directly used to test convergence as the other estimators require knowledge that is not readily available from observations. However, as we have seen, if the convergence happens for E3, then it happens for the other estimators as well.}, converge. We have found that for reasonable choices of parameters -- mainly the smoothing length $S$ -- convergence can be achieved quickly, often after $2\sim4$ iterations. The final results are fairly insensitive to the galaxy bias parameters after three iterations, but depend more strongly on the smoothing parameters $S, C_{\rm ani}$ and the linear growth rate $f$ (particularly when $f$ is too small). With our optimal choice of $S, C_{\rm ani}, b^{(k)}$, $f$, we find that the method can \revise{accurately eliminate the quadrupole moment of the correlation function of the reconstructed galaxy catalogue and reproduce the monopole at $s\gtrsim20\Mpch$}. One thing worthwhile to note here is that the elimination of galaxy correlation quadrupole by tuning $S$ and $C_{\rm ani}$ can be done on real observational data without the use of mock catalogues, which means that the method (with appropriate generalisation to include effects such as survey geometry and completeness) can be applied directly to real data. This can be considered as some sort of {\it self} or {\it internal} calibration of reconstruction parameters, which may help us to avoid relying too heavily on mock data for guiding the choice of parameters.

\begin{table*}
    \centering
    \caption{ {The correlation coefficients before and after the reconstruction procedure. We present the correlation between galaxy and matter distribution, and their combinations. The correlations are charaterized in terms of $k_{90}$ and $k_{50}$, which are the $k$ values where the correlation coefficient falls to $90\%$ and $50\%$, respectively. $k_{50}$ and $k_{90}$ are quoted in unit of $h$~Mpc$^{-1}$. The results shown here are found by reading the corresponding values from  Fig.~\ref{fig:Cor_real_B}.}
    }
    \renewcommand{\arraystretch}{1.2}
    \begin{tabular}{ @{} p{12cm} c c c @{} } 
        \hline\hline
        Correlation coefficient between & Notation & $k_{90}$ &  $k_{50}$ \\
        \hline  
        
        \multicolumn{3}{c}{\bf real-space galaxy distribution and} \\
        
        redshift-space galaxy distribution & $r\left[\delta^s_{\rm g},\delta^r_{\rm g}\right]$ & $0.23$ & $0.68$ \\  
        reconstructed real-space galaxy distribution (starting from redshift space) & $r\left[\delta^{(3)}_{\rm g},\delta^r_{\rm g}\right]$ & $0.32$ & $1.02$ \\
        \\
        
        \multicolumn{3}{c}{\bf initial matter density field and} \\
        real-space galaxy distribution & $r\left[\delta^r_{\rm g},\delta_{\rm ini}\right]$ & $0.08$ & $0.19$ \\
        redshift-space galaxy distribution & $r\left[\delta^s_{\rm g},\delta_{\rm ini}\right]$ & $0.07$ & $0.17$ \\[.4cm]
        
        reconstructed linear matter density from real-space galaxy distribution  & $r\left[\delta_{\rm rec},\delta_{\rm ini}\right]$ & $0.12$ & $0.35$ \\  
        reconstructed linear matter density from redshift-space galaxy distribution (without iterative RSD removal) & $r\left[\delta^{(0)}_{\rm rec},\delta_{\rm ini}\right]$  & $0.11$ & $0.30$ \\ 
        reconstructed linear matter density from redshift-space galaxy distribution (with iterative RSD removal) & $r\left[\delta^{(3)}_{\rm rec},\delta_{\rm ini}\right]$  & $0.12$ & $0.31$ \\
        \hline\hline
    \end{tabular}
    \renewcommand{\arraystretch}{1.0}
    \label{tab:correlation_coefficients}
\end{table*}


{We can quantify the resulting improvement of the reconstruction method by measuring $k_{90}$ and $k_{50}$ which are, respectively, the values of $k$ where the correlation coefficient between two distribution drops below $90\%$ and $50\%$. In Table~\ref{tab:correlation_coefficients} we summarise the values read from Fig.~\ref{fig:Cor_real_B}. The result of $k_{50}$ shows that our iterative RSD removal method increases the $k$-range which has a good (i.e., higher than $50\%$) correlation between the reconstructed and the true real-space galaxy positions by $50\%$. The improvement is even stronger when estimating the initial conditions, with an $80\%$ increase of $k_{50}$ in the correlation between the reconstructed matter density and the true linear density field, compared with the case in which no reconstruction has been applied, if the reconstruction is performed on the redshift-space galaxy distribution. This is similar to, albeit slightly lower than, the case when the initial condition is reconstructed from the real-space galaxy distribution; in this latter case $k_{50}$ increases by a factor of $2$. This confirms that redshift-space distortions have a negative impact that can not be completely undone by this iterative RSD removal method, but the latter leads to a substantial gain.}

The galaxy catalogue we used in this work for tests has a relatively low number density, $n_{\rm g}\simeq3.2\times10^{-4}\left[\Mpch\right]^{-3}$, as a result of which the RSD effect on the reconstruction of initial matter density field is small (e.g., by comparing the dashed grey lines in the right panel of Fig.~\ref{fig:Cor_real_B} or Fig.~\ref{fig:Cor_real_F}). Nevertheless, correcting for RSD still improves the reconstruction result, by increasing the correlation of the reconstructed density field with the true initial density field of our test simulation. It will be interesting to analyse how the method works for galaxy catalogues with higher number density or with other tracers (such as 21cm intensity maps or quasars), in particular whether different choices of parameters need to be made in those cases; we will leave a more detailed investigation of these issues into a future publication.

{The observations that the quadrupole moment can be successfully brought back to $0.0$, and the monopole close to the real-space galaxy correlation function, by the iterative reconstruction method at $s\gtrsim20\Mpch$, raise the interesting question whether the method here has done more than removing linear RSD on these scales. This question would be best answered by quantifying how the constraints on cosmological parameters can be improved by going from linear RSD modelling to analysing summary statistics that are extracted from a reconstructed galaxy catalogue. This is beyond the scope of this work, not least because the current pipeline still needs further extensions to account for varius observational systematics (as mentioned below), but also because future work is needed to understand the covariance after reconstruction (which is nontrivial given that cosmological parameters are used in both the reconstruction-based RSD removal itself and the theoretical modelling of said summary statistics). Without such an analysis available, a somehow indirect way to gain some insight into this question is to look at the scale at which nonlinearity already needs to be accounted for when theoretically predicting the redshift-space correlation function quadrupole. As an exapmle, \citet[][Fig.~14]{Cuesta-Lazaro2020} compared the quadrupoles measured from N-body simulations and from a Gaussian streaming model where the real-space correlation function and galaxy pairwise velocity moments were obtained using the convolutional Lagrangian perturbation theory, and found the latter deviates from simulation results already at scales as large as $40$-$50h^{-1}$Mpc. We also know that the linear Kaiser model does not match simulation data well at scales $\lesssim40$-$50\Mpch$ \citep[see, e.g., the right panel of Fig.~5 in][which used the same set of GR simulations as in this paper]{Hernandez-Aguayo:2018oxg}. These imply that the iterative method does more than linear RSD removal -- which is not surprising given that, unlike the standard Zel'dovich approximation, here the reconstruction method gives a nonlinear displacement field. 
}

The results also imply that it might be possible to use this reconstruction method to infer statistical information about the large-scale peculiar velocity field. The exact details of this information, the ranges of scale and velocity within which it can be reliably extracted, as well as the accuracy of the results, are again beyond the scope of a single paper. Velocity reconstruction has been studied by several groups \citep[e.g.,][]{Wang2011,Yu2019Vel} in a similar context, but with different reconstruction algorithms, including the one used here but with no iterations. It will be of interest to test the iterative reconstruction method presented in this paper for galaxy catalogues with different number densities and at different redshifts in a future work.

We checked the possibility of using this reconstruction method to determine the value of $\beta$, and found that the best-fit $\beta$ from the estimator $R(s)=\xi_0(s)/\xi_{0}\left[{\bf x}^{(k)}\right](s)$ follows the $\beta$ corresponding to the input values of $f, b$ in the reconstruction, and therefore $R(s)$ alone cannot fix $\beta$. However, demanding that the average value of $\xi_2\left[{\bf x}^{(k)}\right](s)$ on large scales, $\bar{\xi}_2\left[{\bf x}^{(k)}\right]$, vanishes could be an alternative way to fix $\beta$. We also discussed briefly the possibility of using a hybrid approach to break the degeneracy in $\beta$ and determine both $f$ and $b$. In this sense the reconstruction can simultaneously give us the cosmological parameter values, the real space galaxy catalogue, the large-scale peculiar velocity field and the initial matter density field. Alternatively, one can use the standard methods to determine $f$, $b$, and use such values to do reconstruction, which still gives the other quantities and this has the advantage that these quantities are for our particular realisation of the Universe. 

The removal of peculiar-velocity-induced modifications to the real galaxy coordinates means that it is possible to get rid of the need to model the effect of RSD on observables, as we have seen above with the example of galaxy two-point correlation function. The latter is indeed one of the most well-understood probes of galaxy clustering, and there are other probes for which the modelling of RSD effects is less widely-studied. Some examples are the higher-order correlation functions of galaxies \citep[e.g.,][]{Slepian:2016weg} and the cross correlation of galaxies with clusters \citep[e.g.,][]{Zu:2013} or voids \citep[e.g.,][]{Hamaus:2015yza,Cai:2016jek,Nadathur:2017jos}. These are situations where reconstruction can be useful by producing reconstructed galaxy catalogues from which the corresponding real-space quantities can be measured; this also has the advantage that the same reconstructed galaxy catalogue can be used to study different probes, rather than having to build a RSD model for each of them. We note that reconstruction-based method has recently been applied to study voids by \citet{Nadathur:2018pjn}. We also note that, even for the redshift-space two-point galaxy correlation function, obtaining percent-level accuracy in the theoretical predictions down to $s\sim20\Mpch$ is challenging \citep[see, e.g.,][for some recent efforts to improve the modelling]{Bianchi:2014kba,Bianchi:2016qen,Kuruvilla:2017kev}.

Another field where reconstruction can be helpful is the study of dark energy and modified gravity models \citep[see, e.g.,][for some latest reviews]{Baker:2019gxo,MGbook}. In these models large-scale structure formation can be more complicated than in $\Lambda$CDM, making it more challenging to predict RSD effects \citep[e.g.,][]{Bose:2016qun,Valogiannis:2019nfz}. The removal of such effects through reconstruction can help to simplify the development of theoretical templates for constraining these models \citep[e.g.,][]{Koyama:2009xx,Aviles:2017aor}. However, we caution that in the current implementation of our reconstruction algorithm we have used the assumption that linear bias is scale-independent on large scales. While this assumption is a good one for some models, it may break down for others for which a more detailed study is needed to understand how best to incorporate their nontrivial bias behaviour. We note that including galaxy biases beyond linear order has been studied in \citet{Birkin:2018nag} which demonstrated its feasibility.

Finally, in this work we have studied the reconstruction in `idealised' galaxy catalogues in periodic cubic simulation boxes, while observational systematics such as survey geometry, selection function and redshift failure are not included in the analysis. These must be tested carefully using more realistic mock catalogues where such systematics exist \citep[e.g.,][]{Smith:2017tzz}, such as done by \citet{Hada:2018ziy}. One advantage of reconstruction in real space is that survey boundary effects are localised and the induced error is primarily restricted to regions near the boundaries \citep{Mao2019}, and we expect the same will apply to our reconstruction algorithm as well. We also expect the effect of redshift evolution on galaxy clustering to be small in our reconstruction method, as long as the redshift dependence of galaxy bias is properly accounted for. Another complication not taken into account in the analysis here is that, even without peculiar velocities, translating a galaxy's redshift to its radial coordinate requires the assumption of a fiducial cosmological model -- an incorrect one would lead to anistropies in the galaxy correlation function even before RSD effects are included, and these anisotropies could be confused as a RSD signal, leading to incorrect reconstruction outcomes. In the standard approach, it is possible to transform the theoretical prediction of the redshift space correlation function to the fiducial cosmology \citep[e.g.,][]{Sanchez:2016sas} used to infer the correlation function from galaxy redshifts. We expect the hybrid approach mentioned in the end of Section \ref{sect:application} (for which a cosmological model also has to be assumed) will help to break this degeneracy in a similar way, but a detailed investigation of this will be left for future work.

\section*{Acknowledgements}

We thank Shaun Cole and Jie Wang for useful comments, and Jiamin Hou for early discussions on this subject. BL is supported by the European Research Council via an ERC Starting Grant (ERC-StG-716532-PUNCA) and the UK Science and Technology Facilities Council (STFC) via Consolidated Grant No.~ST/L00075X/1. MC is supported by the EU Horizon 2020 Research and Innovation programme under a Marie Sk{\l}odowska-Curie grant agreement 794474 (DancingGalaxies). This work used the DiRAC Data Centric system at Durham University, operated by the Institute for Computational Cosmology on behalf of the STFC DiRAC HPC Facility (\url{www.dirac.ac.uk}). This equipment was funded by BIS National E-infrastructure capital grant ST/K00042X/1, STFC capital grants ST/H008519/1, ST/K00087X/1, STFC DiRAC Operations grant ST/K003267/1 and Durham University. DiRAC is part of the National E-Infrastructure. The data underlying this article will be shared on reasonable request to the corresponding author.








\appendix




\bibliographystyle{mnras}
\bibliography{references}

\bsp	
\label{lastpage}
\end{document}